\documentclass[preprint2]{aastex63}

\usepackage{graphicx}% Include figure files
\usepackage{dcolumn}% Align table columns on decimal point
\usepackage{bm}% bold math
\usepackage{subfigure}
\usepackage[nointegrals]{wasysym}
\usepackage{verbatim}
\usepackage{color}
\usepackage{amsmath}
\usepackage[utf8]{inputenc}

\newcommand{\bL}{{\boldsymbol{L}}}

\newcommand{\rmn}{\mathrm}
\newcommand{\N}{\mathcal{N}}

\newcommand{\be}{\begin{equation}}
\newcommand{\ee}{\end{equation}}
\newcommand{\ba}{\begin{eqnarray}}
\newcommand{\ea}{\end{eqnarray}}

\newcommand{\barr}{\begin{array}}
\newcommand{\earr}{\end{array}}

\newcommand{\meanmass}{1.7\pm0.4}
\newcommand{\meanz}{1.08}
\newcommand{\nclusters}{677~}
\newcommand{\tnclusters}{1676~}
\newcommand{\curlpte}{0.05}
\newcommand{\curlsigma}{1.9}
\newcommand{\fitpte}{0.93}
\newcommand{\fitsigma}{4.2}
\newcommand{\nullsigma}{4.9}

\def\ellb{{\boldsymbol{ \ell}}}
\def\x{{\bf x}}
\def\r{{\bf r}}
\def\y{{\bf y}}
\def\n{{\bf n}}

\newcommand{\madcows}{MaDCoWS}
\newcommand{\Planck}{{\it Planck}}
\newcommand{\Msun}{\mathrm{M}_\odot}
\newcommand{\observable}{\lambda}

\usepackage{aas_macros}

\begin{document}

\title{The Atacama Cosmology Telescope: \\
Weighing distant clusters with the most ancient light}
 \shorttitle{Weighing distant clusters with the most ancient light}
  \shortauthors{Madhavacheril, Sif\'{o}n, Battaglia et al.}

\correspondingauthor{Mathew~S.~Madhavacheril}
\email{mmadhavacheril@perimeterinstitute.ca}

\correspondingauthor{Crist\'{o}bal~Sif\'{o}n}
\email{cristobal.sifon@pucv.cl}

\author{Mathew~S.~Madhavacheril}
\affiliation{Centre for the Universe, Perimeter Institute, Waterloo, ON N2L 2Y5, Canada}

\author{Crist\'{o}bal~Sif\'{o}n}
\affiliation{Instituto de F\'isica, Pontificia Universidad Cat\'olica de Valpara\'iso, Casilla 4059, Valpara\'iso, Chile}

\author{Nicholas~Battaglia}
\affiliation{Department of Astronomy, Cornell University, Ithaca, NY 14853 USA}

\author{Simone~Aiola}
\affiliation{Center for Computational Astrophysics, Flatiron Institute, 162 5th Avenue, New York, NY, USA 10010}

\author{Stefania~Amodeo}
\affiliation{Department of Astronomy, Cornell University, Ithaca, NY 14853 USA}

\author{Jason~E.~Austermann}
\affiliation{NIST Quantum Devices Group, 325 Broadway Mailcode 817.03, Boulder, CO, USA 80305}

\author{James~A.~Beall}
\affiliation{NIST Quantum Devices Group, 325 Broadway Mailcode 817.03, Boulder, CO, USA 80305}

\author{Daniel~T.~Becker}
\affiliation{NIST Quantum Devices Group, 325 Broadway Mailcode 817.03, Boulder, CO, USA 80305}

\author{J.~Richard~Bond}
\affiliation{Canadian Institute for Theoretical Astrophysics, University of Toronto, 60 St. George Street, Toronto, ON, Canada, M5S 3H8}

\author{Erminia~Calabrese}
\affiliation{School of Physics and Astronomy, Cardiff University, The Parade, Cardiff, CF24 3AA, UK}

\author{Steve~K.~Choi}
\affiliation{Department of Physics, Cornell University, Ithaca, NY 14853 USA}
\affiliation{Department of Astronomy, Cornell University, Ithaca, NY 14853 USA}

\author{Edward~V.~Denison}
\affiliation{NIST Quantum Devices Group, 325 Broadway Mailcode 817.03, Boulder, CO, USA 80305}

\author{Mark~J.~Devlin}
\affiliation{Department of Physics and Astronomy, University of Pennsylvania, 209 South 33rd Street, Philadelphia, PA 19104}

\author{Simon~R.~Dicker}
\affiliation{Department of Physics and Astronomy, University of Pennsylvania, 209 South 33rd Street, Philadelphia, PA 19104}

\author{Shannon~M.~Duff}
\affiliation{NIST Quantum Devices Group, 325 Broadway Mailcode 817.03, Boulder, CO, USA 80305}

\author{Adriaan~J.~Duivenvoorden}
\affiliation{Joseph Henry Laboratories of Physics, Jadwin Hall, Princeton University, Princeton, NJ 08544, USA}

\author{Jo~Dunkley}
\affiliation{Department of Astrophysical Sciences, Princeton University, 4 Ivy Lane, Princeton, NJ, USA 08544}
\affiliation{Joseph Henry Laboratories of Physics, Jadwin Hall, Princeton University, Princeton, NJ 08544, USA}

\author{Rolando~D\"unner}
\affiliation{Instituto de Astrof\'isica and Centro de Astro-Ingenier\'ia, Facultad de F\'isica, Pontificia Universidad Cat\'olica de Chile, Av.  Vicu\~na Mackenna 4860, 7820436 Macul, Santiago, Chile}

\author{Simone~Ferraro}
\affiliation{Physics Division, Lawrence Berkeley National Laboratory, Berkeley, CA 94720, USA}

\author{Patricio~A.~Gallardo}
\affiliation{Department of Physics, Cornell University, Ithaca, NY 14853 USA}

\author{Yilun~Guan}
\affiliation{Department of Physics and Astronomy, University of Pittsburgh, Pittsburgh, PA, USA 15260}

\author{Dongwon~Han}
\affiliation{Physics and Astronomy Department, Stony Brook University, Stony Brook, NY 11794, USA}

\author{J.~Colin~Hill}
\affiliation{Department of Physics, Columbia University, 550 West 120th Street, New York, NY, USA 10027}
\affiliation{Center for Computational Astrophysics, Flatiron Institute, 162 5th Avenue, New York, NY, USA 10010}

\author{Gene~C.~Hilton}
\affiliation{NIST Quantum Devices Group, 325 Broadway Mailcode 817.03, Boulder, CO, USA 80305}

\author{Matt~Hilton}
\affiliation{Astrophysics Research Centre, University of KwaZulu-Natal, Westville Campus, Durban 4041, South Africa}
\affiliation{School of Mathematics, Statistics \& Computer Science, University of KwaZulu-Natal, Westville Campus, Durban 4041, South Africa}

\author{Johannes~Hubmayr}
\affiliation{NIST Quantum Devices Group, 325 Broadway Mailcode 817.03, Boulder, CO, USA 80305}

\author{Kevin~M.~Huffenberger}
\affiliation{Department of Physics, Florida State University, Tallahassee, FL 32306, USA}

\author{John~P.~Hughes}
\affiliation{Department of Physics and Astronomy, Rutgers University, 136 Frelinghuysen Road, Piscataway, NJ 08854-8019 USA}

\author{Brian~J.~Koopman}
\affiliation{Department of Physics, Yale University, New Haven, CT 06520, USA}

\author{Arthur~Kosowsky}
\affiliation{Department of Physics and Astronomy, University of Pittsburgh, Pittsburgh, PA, USA 15260}

\author{Jeff~Van~Lanen}
\affiliation{NIST Quantum Devices Group, 325 Broadway Mailcode 817.03, Boulder, CO, USA 80305}

\author{Eunseong~Lee}
\affiliation{Jodrell Bank Centre for Astrophysics, School of Physics and Astronomy, University of Manchester, Manchester, UK}

\author{Thibaut~Louis}
\affiliation{Universit\'e Paris-Saclay, CNRS/IN2P3, IJCLab, 91405 Orsay, France}

\author{Amanda~MacInnis}
\affiliation{Physics and Astronomy Department, Stony Brook University, Stony Brook, NY 11794, USA}

\author{Jeffrey~McMahon}
\affiliation{Kavli Institute for Cosmological Physics, University of Chicago, 5640 S. Ellis Ave., Chicago, IL 60637, USA}
\affiliation{Department of Astronomy and Astrophysics, University of Chicago, 5640 S. Ellis Ave., Chicago, IL 60637, USA}
\affiliation{Department of Physics, University of Chicago, Chicago, IL 60637, USA}
\affiliation{Enrico Fermi Institute, University of Chicago, Chicago, IL 60637, USA}

\author{Kavilan~Moodley}
\affiliation{Astrophysics Research Centre, University of KwaZulu-Natal, Westville Campus, Durban 4041, South Africa}
\affiliation{School of Mathematics, Statistics \& Computer Science, University of KwaZulu-Natal, Westville Campus, Durban 4041, South Africa}

\author{Sigurd~Naess}
\affiliation{Center for Computational Astrophysics, Flatiron Institute, 162 5th Avenue, New York, NY, USA 10010}

\author{Toshiya~Namikawa}
\affiliation{Center for Theoretical Cosmology, DAMTP, University of Cambridge, CB3 0WA, UK}

\author{Federico~Nati}
\affiliation{Department of Physics, University of Milano-Bicocca, Piazza della Scienza 3, 20126 Milano, Italy}

\author{Laura~Newburgh}
\affiliation{Department of Physics, Yale University, New Haven, CT 06520, USA}

\author{Michael~D.~Niemack}
\affiliation{Department of Physics, Cornell University, Ithaca, NY 14853 USA}
\affiliation{Department of Astronomy, Cornell University, Ithaca, NY 14853 USA}

\author{Lyman~A.~Page}
\affiliation{Joseph Henry Laboratories of Physics, Jadwin Hall, Princeton University, Princeton, NJ 08544, USA}

\author{Bruce~Partridge}
\affiliation{Department of Physics and Astronomy, Haverford College,Haverford, PA, USA 19041}

\author{Frank~J.~Qu}
\affiliation{Center for Theoretical Cosmology, DAMTP, University of Cambridge, CB3 0WA, UK}

\author{Naomi~C.~Robertson}
\affiliation{Institute of Astronomy, Madingley Road, Cambridge CB3 0HA, UK}
\affiliation{Kavli Institute for Cosmology, University of Cambridge, Madingley Road, Cambridge CB3 OHA, UK}

\author{Maria~Salatino}
\affiliation{Physics Department, Stanford University, 382 via Pueblo, Stanford, CA 94305, USA}
\affiliation{Kavli Institute for Particle Astrophysics and Cosmology, 452 Lomita Mall, Stanford, CA 94305-4085, USA}

\author{Emmanuel~Schaan}
\affiliation{Physics Division, Lawrence Berkeley National Laboratory, Berkeley, CA 94720, USA}

\author{Alessandro~Schillaci}
\affiliation{Department of Physics, California Institute of Technology, Pasadena, CA 91125, USA}

\author{Benjamin~L.~Schmitt}
\affiliation{Harvard-Smithsonian Center for Astrophysics, Harvard University, 60 Garden St, Cambridge, MA 02138, United States}

\author{Neelima~Sehgal}
\affiliation{Physics and Astronomy Department, Stony Brook University, Stony Brook, NY 11794, USA}

\author{Blake~D.~Sherwin}
\affiliation{Center for Theoretical Cosmology, DAMTP, University of Cambridge, CB3 0WA, UK}
\affiliation{Kavli Institute for Cosmology, University of Cambridge, Madingley Road, Cambridge CB3 OHA, UK}

\author{Sara~M.~Simon}
\affiliation{Fermi National Accelerator Laboratory, Batavia, IL 60510, USA}

\author{David~N.~Spergel}
\affiliation{Center for Computational Astrophysics, Flatiron Institute, 162 5th Avenue, New York, NY, USA 10010}
\affiliation{Department of Astrophysical Sciences, Princeton University, 4 Ivy Lane, Princeton, NJ, USA 08544}

\author{Suzanne~Staggs}
\affiliation{Joseph Henry Laboratories of Physics, Jadwin Hall, Princeton University, Princeton, NJ 08544, USA}

\author{Emilie~R.~Storer}
\affiliation{Joseph Henry Laboratories of Physics, Jadwin Hall, Princeton University, Princeton, NJ 08544, USA}

\author{Joel~N.~Ullom}
\affiliation{NIST Quantum Devices Group, 325 Broadway Mailcode 817.03, Boulder, CO, USA 80305}

\author{Leila~R.~Vale}
\affiliation{NIST Quantum Devices Group, 325 Broadway Mailcode 817.03, Boulder, CO, USA 80305}

\author{Alexander~van~Engelen}
\affiliation{School of Earth and Space Exploration and Department of Physics, Arizona State University, Tempe, AZ 85287}

\author{Eve~M.~Vavagiakis}
\affiliation{Department of Physics, Cornell University, Ithaca, NY 14853 USA}

\author{Edward~J.~Wollack}
\affiliation{NASA/Goddard Space Flight Center, Greenbelt, MD 20771, USA}

\author{Zhilei~Xu}
\affiliation{Department of Physics and Astronomy, University of Pennsylvania, 209 South 33rd Street, Philadelphia, PA 19104}

\begin{abstract}
    
We use gravitational lensing of the cosmic microwave background (CMB) to measure the mass of the
most distant blindly-selected sample of galaxy clusters on which a lensing measurement has been performed to date.  In CMB data from the the Atacama Cosmology Telescope (ACT) and
the {\it Planck} satellite, we detect the stacked lensing effect from \nclusters near-infrared-selected galaxy clusters from the Massive
and Distant Clusters of WISE Survey (MaDCoWS), which have a mean redshift of $ \langle z \rangle = \meanz$.
There are currently no representative optical weak lensing measurements of clusters that match the distance and average mass of this sample. 
We detect the lensing signal with a significance of $\fitsigma \sigma$.
We model the signal with a halo model framework to  find the mean mass of the population from which these clusters are drawn. Assuming that the clusters
follow Navarro-Frenk-White density profiles, we infer a mean mass of $\langle M_{500c}\rangle = \left(\meanmass\right)\times10^{14}\,\Msun$.  We consider systematic uncertainties from cluster redshift errors, centering errors, and the shape of the NFW profile.  These are all smaller than 30\% of our reported uncertainty. This work highlights the potential of CMB lensing to enable cosmological constraints from the abundance of distant clusters populating ever larger volumes of the observable Universe, beyond the capabilities of optical weak lensing measurements. \\ \\ \\ \\

\end{abstract}

\section{Introduction}

Most of the mass in the Universe is thought to consist of `dark matter' that does not interact with electromagnetic radiation other than through the gravitational force. Distortions in background light sources due to the gravitational influence of massive clusters of galaxies can be used to produce maps of the total matter distribution. This technique has served not only as evidence for the existence of dark matter~\citep{TrimbleReview,MasseyReview}, but also as a method for inferring the total mass of galaxy clusters themselves~\citep[e.g.,][]{HoekstraReview}. Such mass measurements are critical for the program of using the abundance of clusters to infer cosmological parameters like the dark energy equation of state or the mass scale of neutrinos~\citep[e.g.,][]{AEM2011,madhavacheril17}. One key aspect of this program is our ability to constrain the abundance of clusters to high redshifts, directly probing the growth of structure through cosmic time~\citep{Voit2005}.

Sensitive large-area optical/near-infrared surveys are now allowing measurements of the mean mass of clusters up to $z\sim1$ \citep{chiu19,murata19} through the lensing effects induced on background galaxies. However, as the distance of the clusters increases, the number of background galaxies that are useful for weak-lensing measurements decreases rapidly. Measurements of this ``galaxy weak lensing'' effect at large distances are therefore only possible at present through deep targeted observations  with the Hubble Space Telescope \citep{jee11,schrabback18}, with which observations currently exist for only some dozens of rich, 
massive clusters at redshifts of $z>0.8$.
A valuable complementary probe is emerging as cosmic microwave background (CMB) measurements are becoming sensitive enough to allow measurements of weak lensing by galaxy clusters in maps of the temperature and polarization of the CMB \cite[e.g.,][]{Mat2015,Baxter2015,PlnkSZCos2015,GP2017,2019ApJ...872..170R,2019MNRAS.489..401Z}. The high source redshift of the CMB allows weak lensing measurements to higher redshifts than galaxy lensing. Consequently, measurements of CMB lensing by clusters are anticipated to provide more stringent constraints on the masses of high-redshift clusters than enabled by future optical surveys~\citep[e.g.][]{madhavacheril17}.

We provide a mass estimate using gravitational lensing of the CMB for a blindly-selected sample of galaxy clusters whose average mass has not been previously determined using galaxy lensing, which is additionally the highest redshift, $\langle z\rangle=\meanz$, where a detection of gravitational lensing by galaxy clusters has been reported for a blindly-selected sample to date. As opposed to targeted measurements of the most massive clusters, our work allows for inference of the average mass of a representative cluster sample. We make the code used in this analysis available at \url{https://github.com/ACTCollaboration/madcows_lensing}.

\begin{figure}[t]
    \centering
\includegraphics[width=0.9\linewidth]{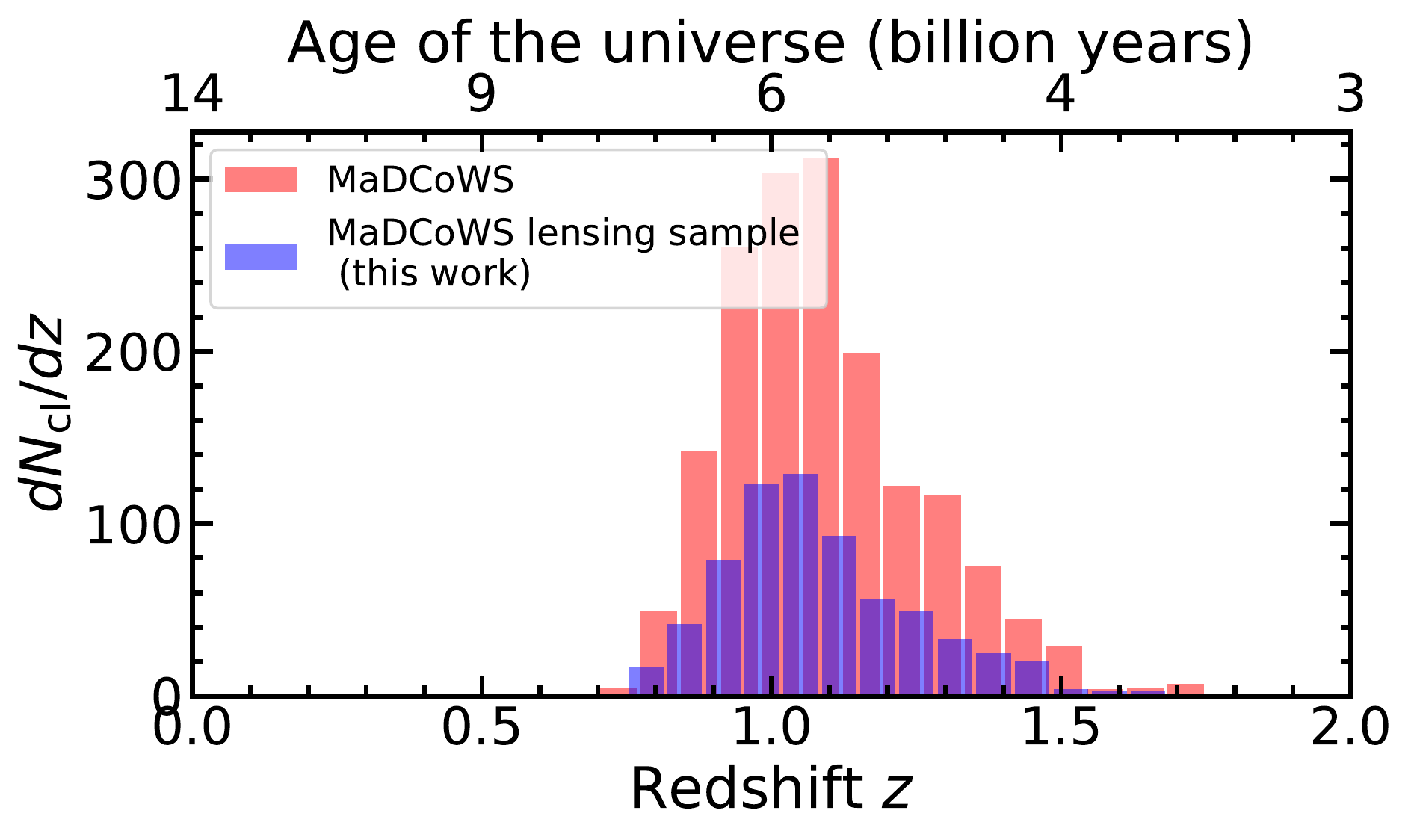}
    \includegraphics[width=0.9\linewidth]{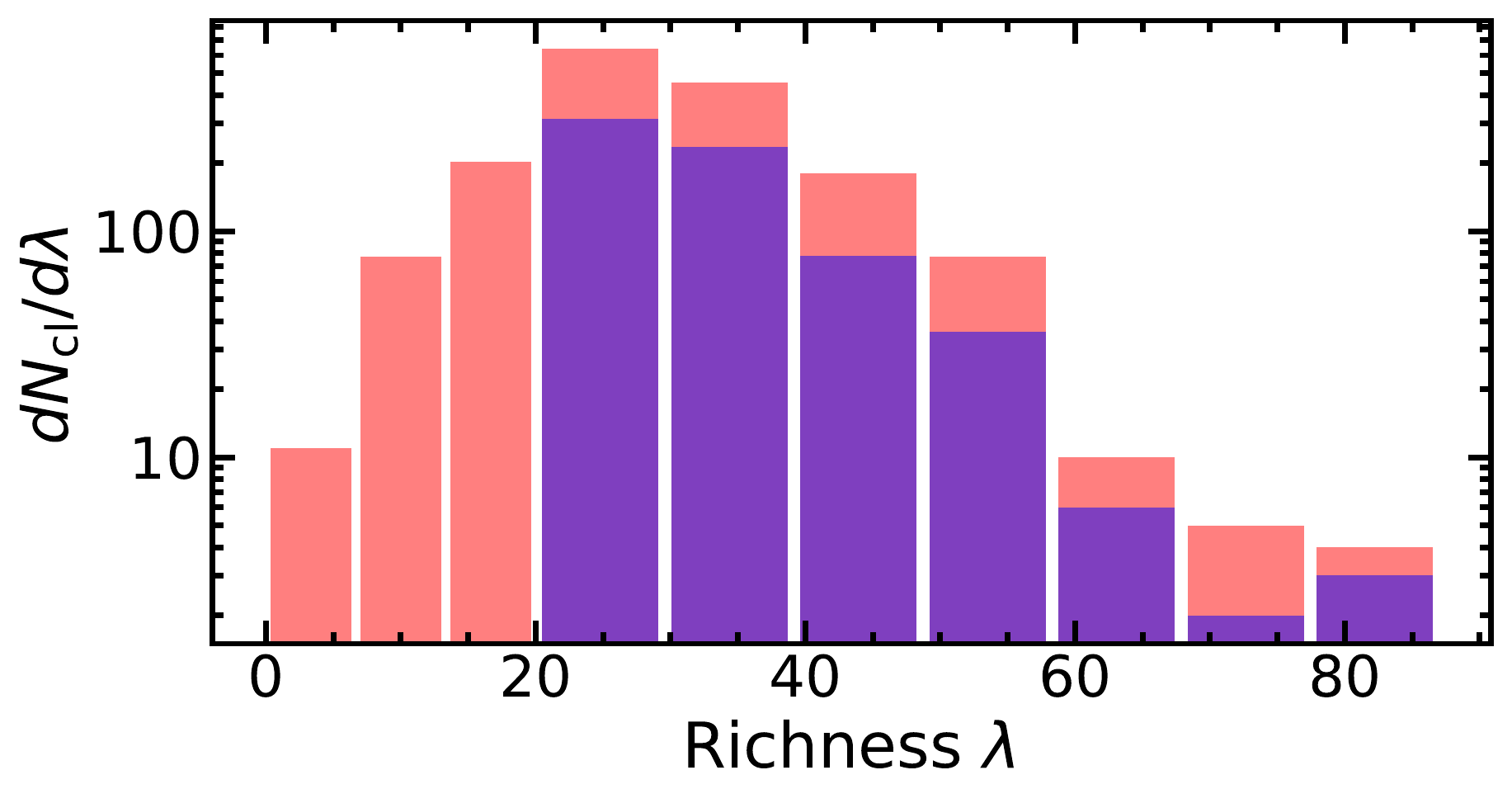}
    \caption{Distributions for cluster redshift (top), and for a measure of the number of galaxies in a cluster, richness (bottom) for the \madcows\ WISE-PanSTARRS galaxy cluster sample used in this work. The blue histograms correspond to the \nclusters clusters that remain after applying the richness cut ($\lambda>20$) and ACT mask, while the red histograms correspond to the full sample of \tnclusters clusters.}
    \label{fig:nz}
\end{figure}

\section{Data}

We use a combination of CMB data from the ground-based Atacama Cosmology Telescope (ACT) and the \Planck\ satellite at the location of galaxy clusters selected from the Massive and Distant Clusters of WISE Survey \citep[\madcows,][]{gonzalez19}.
The \madcows\ clusters were identified as galaxy overdensities in near-infrared imaging (at 3.4~$\mu$m and 4.6~$\mu$m) from the WISE all-sky survey \citep{WISE2010}, and a large number of them were followed up with the \textit{Spitzer} Space Telescope (at 3.6~$\mu$m and 4.5~$\mu$m).
At declinations $>-30^\circ$, the addition of optical data ($grizy$ bands) from the Panoramic Survey Telescope and Rapid Response System \citep[Pan-STARRS,][]{panstarrs} allowed reliable photometric redshift estimation and subsequent cluster richness measurements.
As a first attempt at a mass proxy, \citet{gonzalez19} define the richness, $\observable$, to be the overdensity of red-sequence galaxies (identified using both PanSTARRS and \textit{Spitzer} data) brighter than 15 $\mu$Jy in the \textit{Spitzer} 4.5~$\mu$m band and within 1 Mpc of the brightest cluster galaxy.
Photometric redshifts were estimated from the \textit{Spitzer} 3.6~$\mu$m and 4.5~$\mu$m bands, aided with the PanSTARRS $i$-band to remove low-redshift galaxies, with an estimated scatter $\sigma_z/(1+z)=0.04$ and no significant bias (but with an outlier fraction potentially of order 5\%)\footnote{These statistics are based on a comparison of photometric and spectroscopic redshifts for 38 clusters for which the latter is available.}.
We discuss the impact of photometric redshift uncertainties below.
We use this subset of the \madcows\ ``WISE-PanSTARRS'' sample with available photometric redshifts, imposing an additional $\observable>20$ cut to reduce contamination by false detections in the \madcows\ catalog, for a total of \nclusters clusters after masking point sources and other artifacts in the ACT maps (which is also limited to declinations \ $\leq20^\circ$). 
We show the richness and redshift distributions of these clusters in Fig.\ \ref{fig:nz}. The mean redshift of the sample is $\langle z\rangle=\meanz$ with 
16\% and 84\% percentile range of 0.93--1.26.

To reconstruct the lensing signal, we use co-added maps of ACT and \Planck\ CMB temperature data prepared separately at 98 GHz and 150 GHz and described in \cite{2007.07290}. The co-added maps include night-time data collected during the years 2008 -- 2018 using the MBAC~\citep{Swetz2011}, ACTPol~\citep{thornton/2016} and AdvACT~\citep{Henderson2016} receivers. The \Planck\ maps used in the co-added maps are the PR2 (2015) CMB temperature maps at 100 GHz and 143 GHz. We also use the \Planck\ 2018 SMICA tSZ-deprojected maps~\citep{Planck2018compsep} as an additional input to the lensing reconstruction (see Appendix A). This is done in order to remove the bias from the thermal Sunyaev-Zeldovich (tSZ) effect due to inverse Compton scattering of CMB photons off ionized electrons in hot gas in massive clusters~\citep{1802.08230}.  Deprojection of the tSZ is possible because the frequency dependence of the spectral distortion due to the tSZ effect is well understood.

\begin{figure}[t]
    \centering
    \includegraphics[width=0.95\columnwidth]{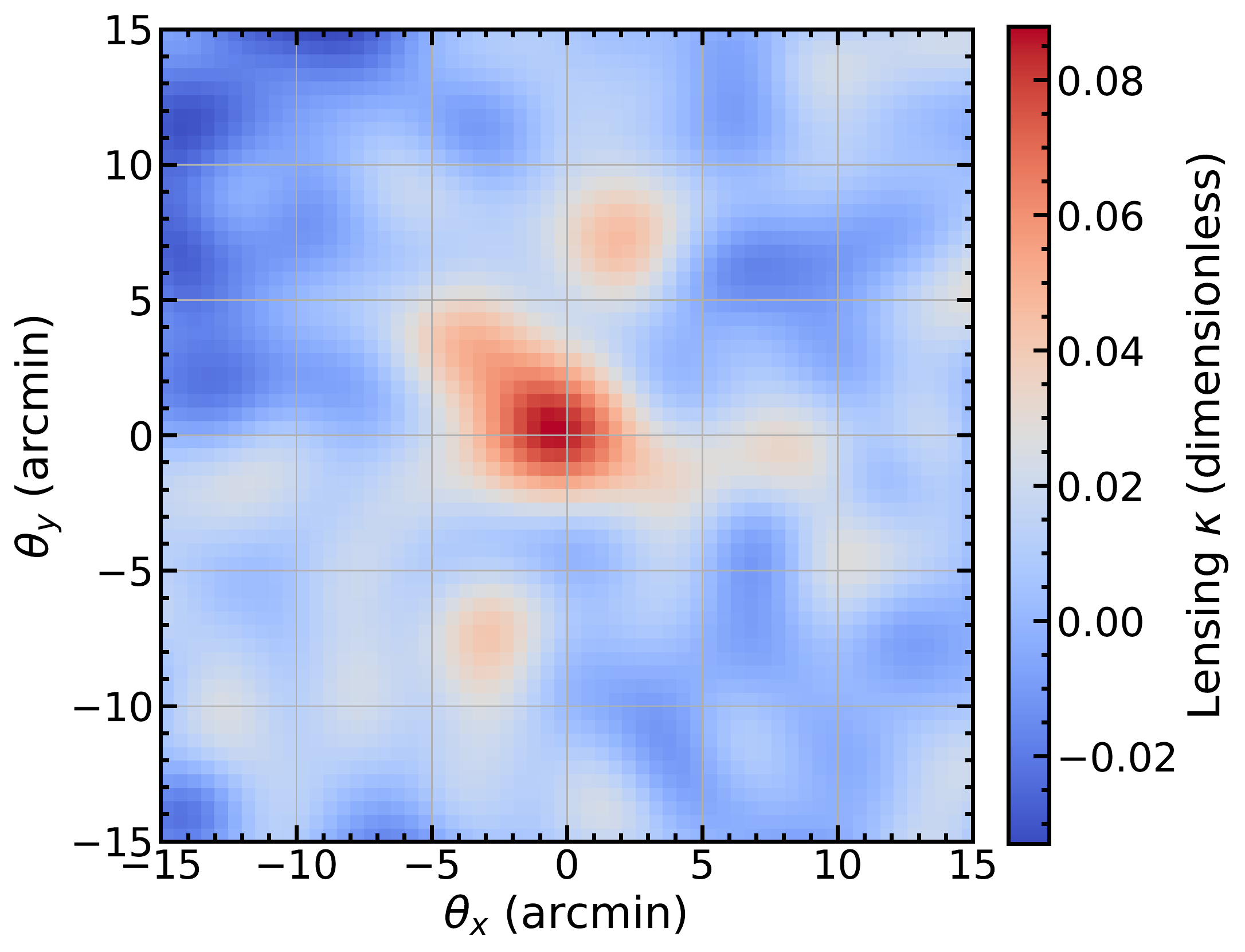}
    \caption{The stacked CMB lensing convergence mass map from the sample of \nclusters MaDCoWS clusters with $\langle z\rangle=\meanz$ used in this work. This reconstruction includes a low-pass filter up to a scale of 2 arcminutes and has been additionally smoothed with a Gaussian filter with FWHM of 3.5~arcminutes. }
    \label{fig:stamp}
\end{figure}

\begin{figure}[htp]
    \centering
    \includegraphics[width=0.95\columnwidth]{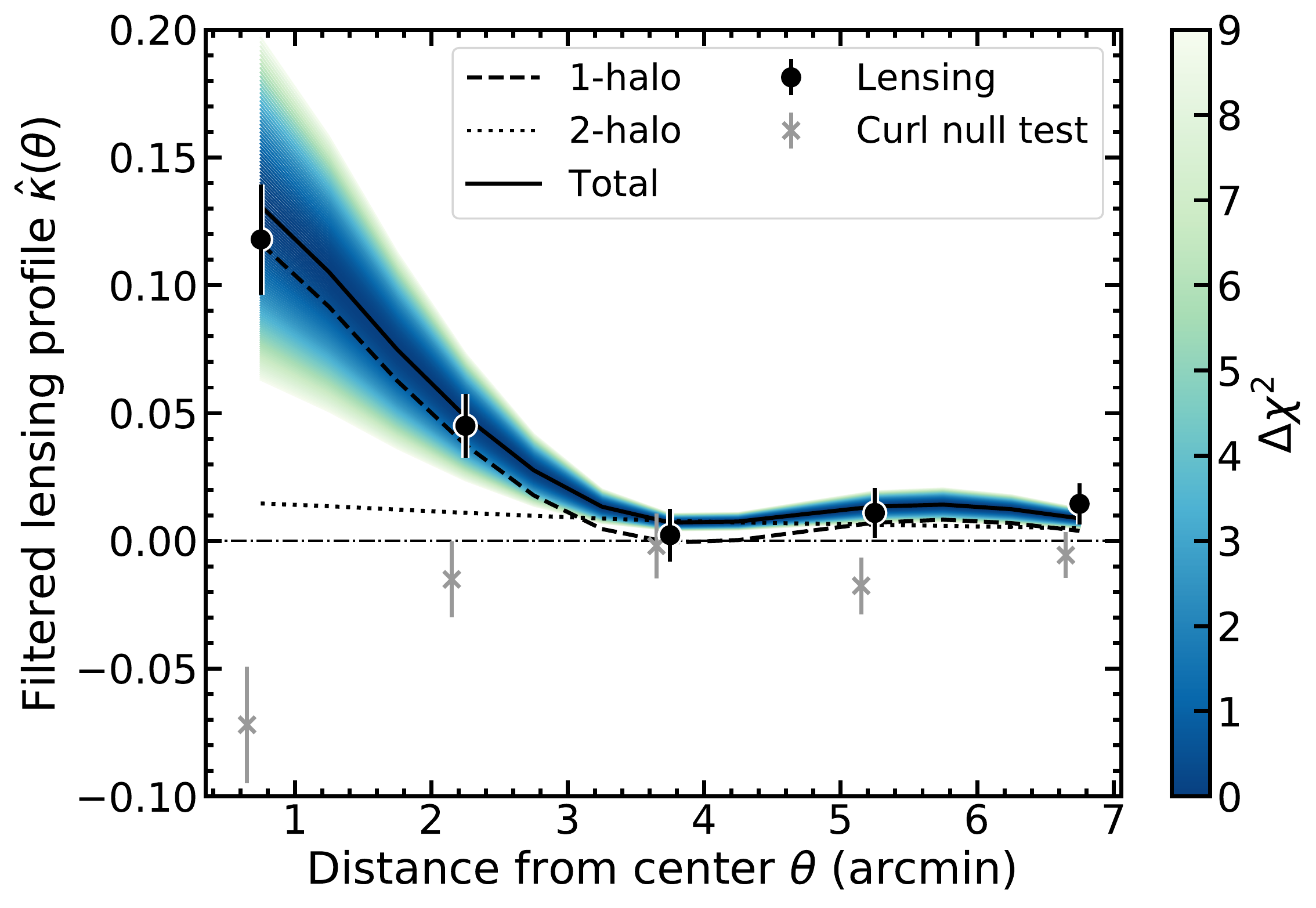}
    \includegraphics[width=0.92\linewidth]{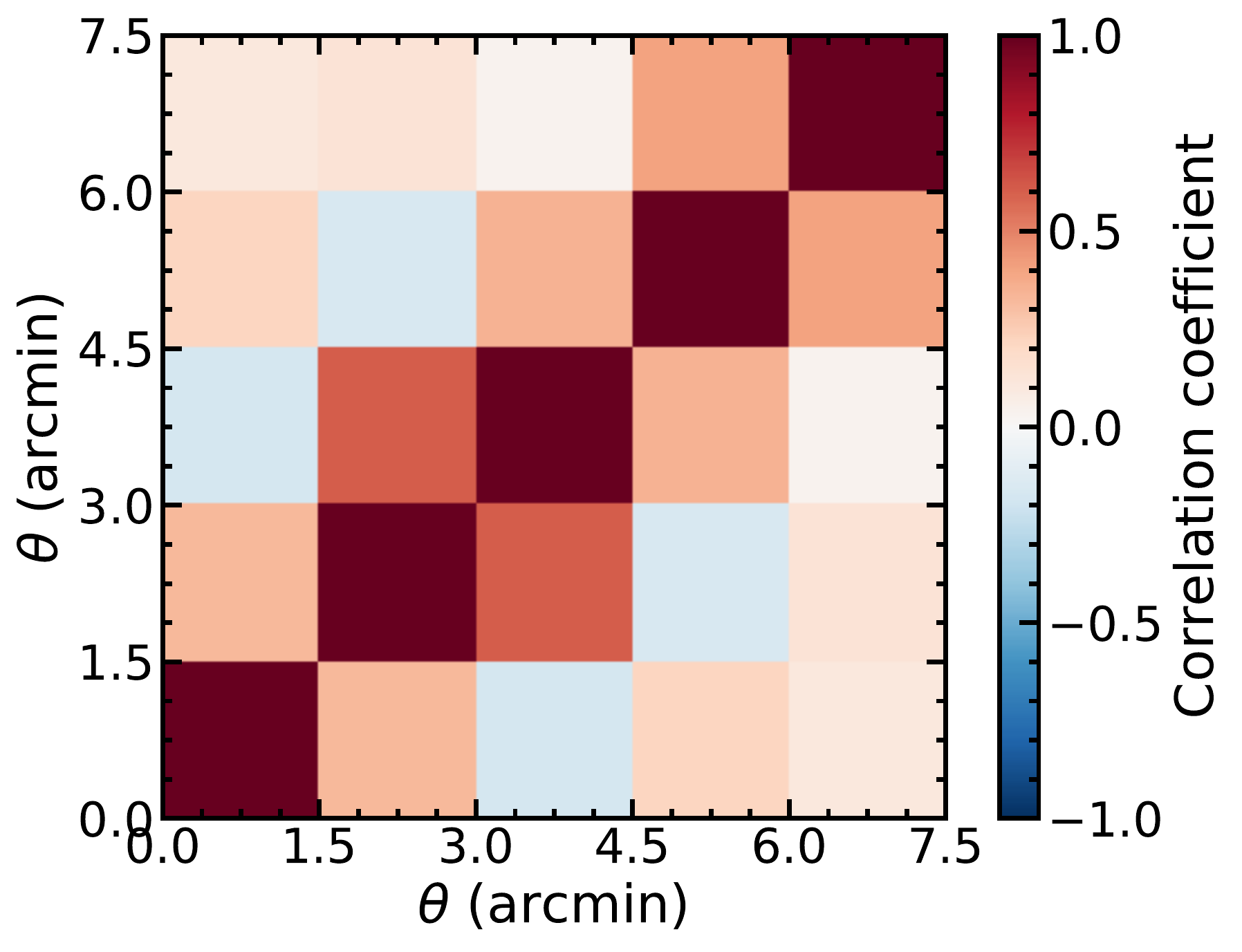}
    \caption{\textit{Top:} Halo model fit to the profile of the stacked lensing convergence, color-coded by $\Delta\chi^2=\chi^2-\chi^2_\mathrm{min}$. With only one free parameter in the model, the 1, 2, and 3$\sigma$ credible regions are limited by $\Delta\chi^2=$1, 4, and 9, respectively. The solid line shows the best-fit model, while the dashed and dot-dashed lines show the contributions from the 1-halo and 2-halo terms to it, respectively. Grey crosses show measurements of the curl (slightly offset horizontally for clarity), which is expected to be zero (see Appendix A). Error-bars correspond to the square root of the diagonal elements of the covariance matrix. The probability-to-exceed (PTE) the $\chi^2$ of our best-fit model is \fitpte. The PTE of the curl compared to a null signal is \curlpte.
    \textit{Bottom:} The correlation coefficient matrix for the radial bins of the lensing measurement. The corresponding matrix for the curl measurement is similar. The large bin-bin correlations show that all the curl points fluctuating below zero is not unlikely.  }
    \label{fig:profile}
\end{figure}

\section{Lensing reconstruction}

We reconstruct the CMB lensing convergence $\kappa$ in regions centered on the locations of \madcows\ clusters. The lensing convergence is related to the line-of-sight integral $\phi$ of the underlying gravitational potential sourced by a cluster, and to the lensing deflection angle $\boldsymbol{\alpha}$ through $\nabla^2 \phi=\boldsymbol{\nabla}\cdot{\boldsymbol{\alpha}}=-2\kappa$. The convergence is related to the surface mass density $\Sigma$ through
$\kappa=\Sigma/\Sigma_{\rm cr}$, where $\Sigma_{\rm cr}$ is a characteristic mass density for the formation of multiple images that depends on distances to the source and lens (see Appendix B). Since the convergence map is directly proportional to the surface mass density, it can be thought of as a mass map. Gravitational lensing of the CMB by clusters leads to a re-mapping of the temperature anisotropies  $T({\bf x}) = T_{\rm unlensed}({\bf x} + \boldsymbol{\alpha})$. In the 2D Fourier space of the temperature anisotropy image, this re-mapping corresponds to coupling between previously independent Fourier modes (say with wavenumbers $\ellb$ and $\ellb'$) that is proportional to the lensing convergence: $\langle T({\boldsymbol \ell})T({\boldsymbol \ell}')  \rangle \propto \kappa(\ellb+\ellb') $. This allows us to reconstruct the underlying convergence mode by mode by using a `quadratic estimator', i.e., a weighted sum over products of pairs of image modes \citep{astro-ph/0701276}, which in practice can be written as the divergence of the product of the large-scale CMB gradient and the small-scale CMB fluctuations (see Appendix A). The quadratic estimator reconstruction provides an unbiased estimate of the Fourier modes of the cluster mass map within a range of scales set by the band-limit of the CMB maps. Thus, the output reconstruction image $\hat{\kappa}({\bf x})$ is effectively filtered, and we forward model this filtering when fitting to theoretical expectations. The signal-to-noise ratio (SNR) for any given cluster is well below unity, and so our final measurements are  a weighted average of the mass maps and of the radial profile (azimuthal average) of the lensing convergence, where the weights are inversely proportional to the noise variance in the lensing convergence reconstruction. We estimate a weighted covariance matrix for the radial profile from the scatter among the profiles. With each cluster profile weighted by $w_i$, the weighted covariance of the weighted mean of the $N$ clusters in our sample is 
\begin{align}
\label{eq:cov}
\mathbf{C} &= \frac{1}{N} \frac {\sum_{i=1}^N w_i \left(\mathbf{p}_i - \mathbf{d}\right)^T \left(\mathbf{p}_i - \mathbf{d}\right)} {V_1 - (V_2 / V_1)}.
\end{align}
where $\mathbf{p}_i$ are the individual radial profiles of each cluster mass map, $\mathbf{d}=\frac{1}{V_1}\sum_{i=1}^N w_i \mathbf{p}_i.
$ is the weighted mean profile, $V_1 = \sum_{i=1}^N w_i$ and $V_2 = \sum_{i=1}^N w_i^2$. Details of the reconstruction process and mitigation of astrophysical foregrounds are provided in Appendix \ref{app:details}. We show the resulting stacked CMB lensing convergence reconstruction in Fig. \ref{fig:stamp}.

\section{Results}

After averaging lensing convergence maps across all the clusters in our sample, we detect an excess relative to null within 8 arcminutes at \nullsigma $\sigma$ confidence. We estimate the mean mass of the sample by fitting the binned radial profile of the average mass map assuming the Navarro-Frenk-White \citep[NFW,][]{NFW} density profile model for the distribution of matter within galaxy clusters with a fixed relation between the profile shape (or `concentration' parameter) and the mass. 
We account for the distribution of clusters of different masses across redshift and their clustering (i.e., the two-halo term), and apply an approximate correction for selection effects in order to construct an ensemble average of the lensing signal. This `halo model' approach, detailed in Appendix \ref{app:model}, allows us to estimate the mean mass of the underlying sample from which the \madcows\ sample is drawn. 

We show the model predictions for the filtered lensing signal in Fig.\ \ref{fig:profile}, color-coded by the excess $\chi^2$ with respect to the minimum $\chi^2$ obtained from our fit, corresponding to $\chi^2_\mathrm{min}=1.4$. Since our model has only one free parameter, the 1, 2, and 3$\sigma$ credible ranges are given by $\Delta\chi^2=1,4,$ and 9, respectively. 
We calculate the probability-to-exceed $\chi^2_\mathrm{min}$ (PTE) by drawing one million random samples with mean zero and using the covariance of our measurements; we find a PTE of \fitpte, showing that the model is an adequate description of the data. The best-fit mean mass is $\langle M_{500c}\rangle =(\meanmass)\times10^{14}\,\Msun$, with a preference for our best-fit over zero mass at the \fitsigma $\sigma$ level.

\section{Discussion}

The clusters in our sample have a mean redshift of $\langle z\rangle=\meanz$, the highest-redshift, blindly-selected sample for which a gravitational lensing measurement has been obtained to date.
Our results highlight the potential of both CMB lensing in general and ACT in particular to constrain the scaling relation between mass and other observables for high-redshift galaxy clusters.

In Appendix B, we explore systematic uncertainties from cluster redshift errors, centering errors, and the concentration parameter of the NFW profile, which are all smaller than 30\% of our reported uncertainty. Since our measurement probes the overall lensing amplitude, the reported mass is also robust to assumptions about both the selection function and the scaling relation between mass and richness, which conversely we are not able to constrain.

Because we stack the lensing maps around all \madcows\ clusters within the ACT footprint with a richness larger than 20, our measurement is representative of the full \madcows\ sample above this cut. Targeted observations of the tSZ effect (which scales steeply with cluster mass as $\sim M^{\frac{5}{3}}$) of \madcows\ sub-samples by \cite{gonzalez19,madcows_aca,madcows_mustang} have naturally picked preferentially the most luminous objects, and thus the mean richness of our sample, $\langle\observable\rangle=32$, is lower than that probed in those works. The mean mass we estimate is consistent with the scaling relations reported in those works, which however have large uncertainties and are susceptible to additional systematic effects associated with an uncertain selection function.

While looking for galaxy overdensities is an efficient means of finding galaxy clusters, characterizing the selection effects that these techniques are subject to is a challenging endeavor. No estimate of the selection function is available for \madcows, which precludes the use of this sample for cosmological parameter inference. Future CMB lensing measurements of well-defined cluster samples at high redshift will yield tight constraints on the parameters that govern the growth of structure. \\

{\it{Acknowledgements:}} We thank Peter Melchior for helpful discussions.
% Software
Software used for this analysis includes healpy~\citep{Healpix1}, HEALPix~\citep{Healpix2}, and Astropy~\citep{astropy:2013, astropy:2018}.
% Funding
Research at Perimeter Institute is supported in part by the Government of Canada through the Department of Innovation, Science and Industry Canada and by the Province of Ontario through the Ministry of Colleges and Universities. CS acknowledges support from the Agencia Nacional de Investigaci\'on y Desarrollo (ANID) through FONDECYT Iniciaci\'on grant no.~11191125. NB acknowledges support from NSF grant AST-1910021. EC acknowledges support from the STFC Ernest Rutherford Fellowship ST/M004856/2 and STFC Consolidated Grant ST/S00033X/1, and from the Horizon 2020 ERC Starting Grant (Grant agreement No 849169).
JD is supported through NSF grant AST-1814971. R.D. thanks CONICYT for grant BASAL CATA AFB-170002. DH, AM, and NS acknowledge support from NSF grant numbers AST-1513618 and AST-1907657.
 MHi acknowledges support from the National Research Foundation.
JPH acknowledges funding for SZ cluster studies from NSF grant number AST-1615657. KM acknowledges support from the National Research Foundation of South Africa. 
% ACT
This work was supported by the U.S. National Science Foundation through awards AST-1440226, AST0965625 and AST-0408698 for the ACT project, as well as awards PHY-1214379 and PHY-0855887. Funding was also provided by Princeton University, the University of Pennsylvania, and a Canada Foundation for Innovation (CFI) award to UBC. ACT operates in the Parque Astron\'{o}mico Atacama in northern Chile under the auspices of the Comisi\'{o}n Nacional de Investigaci\'{o}n Cient\'{i}fica y Tecnol\'{o}gica de Chile (CONICYT). Computations were performed on the GPC and \emph{Niagara} supercomputers at the SciNet HPC Consortium. SciNet is funded by the CFI under the auspices of Compute Canada, the Government of Ontario, the Ontario Research Fund
-- Research Excellence; and the University of Toronto. The development of multichroic detectors and lenses was supported by NASA grants NNX13AE56G and NNX14AB58G.   Colleagues at AstroNorte and RadioSky provide logistical support and keep operations in Chile running smoothly. We also thank the Mishrahi Fund and the Wilkinson Fund for their generous support of the project. This document was prepared by Atacama Cosmology Telescope using the resources of the Fermi National Accelerator Laboratory (Fermilab), a U.S. Department of Energy, Office of Science, HEP User Facility. Fermilab is managed by Fermi Research Alliance, LLC (FRA), acting under Contract No. DE-AC02-07CH11359.

\bibliography{msm}

\begin{thebibliography}{}
\expandafter\ifx\csname natexlab\endcsname\relax\def\natexlab#1{#1}\fi
\providecommand{\url}[1]{\href{#1}{#1}}
\providecommand{\dodoi}[1]{doi:~\href{http://doi.org/#1}{\nolinkurl{#1}}}
\providecommand{\doeprint}[1]{\href{http://ascl.net/#1}{\nolinkurl{http://ascl.net/#1}}}
\providecommand{\doarXiv}[1]{\href{https://arxiv.org/abs/#1}{\nolinkurl{https://arxiv.org/abs/#1}}}

\bibitem[{{Allen} {et~al.}(2011){Allen}, {Evrard}, \& {Mantz}}]{AEM2011}
{Allen}, S.~W., {Evrard}, A.~E., \& {Mantz}, A.~B. 2011, \araa, 49, 409,
  \dodoi{10.1146/annurev-astro-081710-102514}

\bibitem[{{Astropy Collaboration} {et~al.}(2013){Astropy Collaboration},
  {Robitaille}, {Tollerud}, {Greenfield}, {Droettboom}, {Bray}, {Aldcroft},
  {Davis}, {Ginsburg}, {Price-Whelan}, {Kerzendorf}, {Conley}, {Crighton},
  {Barbary}, {Muna}, {Ferguson}, {Grollier}, {Parikh}, {Nair}, {Unther},
  {Deil}, {Woillez}, {Conseil}, {Kramer}, {Turner}, {Singer}, {Fox}, {Weaver},
  {Zabalza}, {Edwards}, {Azalee Bostroem}, {Burke}, {Casey}, {Crawford},
  {Dencheva}, {Ely}, {Jenness}, {Labrie}, {Lim}, {Pierfederici}, {Pontzen},
  {Ptak}, {Refsdal}, {Servillat}, \& {Streicher}}]{astropy:2013}
{Astropy Collaboration}, {Robitaille}, T.~P., {Tollerud}, E.~J., {et~al.} 2013,
  \aap, 558, A33, \dodoi{10.1051/0004-6361/201322068}

\bibitem[{{Astropy Collaboration} {et~al.}(2018){Astropy Collaboration},
  {Price-Whelan}, {Sip{\H{o}}cz}, {G{\"u}nther}, {Lim}, {Crawford}, {Conseil},
  {Shupe}, {Craig}, {Dencheva}, {Ginsburg}, {Vand erPlas}, {Bradley},
  {P{\'e}rez-Su{\'a}rez}, {de Val-Borro}, {Aldcroft}, {Cruz}, {Robitaille},
  {Tollerud}, {Ardelean}, {Babej}, {Bach}, {Bachetti}, {Bakanov}, {Bamford},
  {Barentsen}, {Barmby}, {Baumbach}, {Berry}, {Biscani}, {Boquien}, {Bostroem},
  {Bouma}, {Brammer}, {Bray}, {Breytenbach}, {Buddelmeijer}, {Burke},
  {Calderone}, {Cano Rodr{\'\i}guez}, {Cara}, {Cardoso}, {Cheedella}, {Copin},
  {Corrales}, {Crichton}, {D'Avella}, {Deil}, {Depagne}, {Dietrich}, {Donath},
  {Droettboom}, {Earl}, {Erben}, {Fabbro}, {Ferreira}, {Finethy}, {Fox},
  {Garrison}, {Gibbons}, {Goldstein}, {Gommers}, {Greco}, {Greenfield},
  {Groener}, {Grollier}, {Hagen}, {Hirst}, {Homeier}, {Horton}, {Hosseinzadeh},
  {Hu}, {Hunkeler}, {Ivezi{\'c}}, {Jain}, {Jenness}, {Kanarek}, {Kendrew},
  {Kern}, {Kerzendorf}, {Khvalko}, {King}, {Kirkby}, {Kulkarni}, {Kumar},
  {Lee}, {Lenz}, {Littlefair}, {Ma}, {Macleod}, {Mastropietro}, {McCully},
  {Montagnac}, {Morris}, {Mueller}, {Mumford}, {Muna}, {Murphy}, {Nelson},
  {Nguyen}, {Ninan}, {N{\"o}the}, {Ogaz}, {Oh}, {Parejko}, {Parley}, {Pascual},
  {Patil}, {Patil}, {Plunkett}, {Prochaska}, {Rastogi}, {Reddy Janga},
  {Sabater}, {Sakurikar}, {Seifert}, {Sherbert}, {Sherwood-Taylor}, {Shih},
  {Sick}, {Silbiger}, {Singanamalla}, {Singer}, {Sladen}, {Sooley},
  {Sornarajah}, {Streicher}, {Teuben}, {Thomas}, {Tremblay}, {Turner},
  {Terr{\'o}n}, {van Kerkwijk}, {de la Vega}, {Watkins}, {Weaver}, {Whitmore},
  {Woillez}, {Zabalza}, \& {Astropy Contributors}}]{astropy:2018}
{Astropy Collaboration}, {Price-Whelan}, A.~M., {Sip{\H{o}}cz}, B.~M., {et~al.}
  2018, \aj, 156, 123, \dodoi{10.3847/1538-3881/aabc4f}

\bibitem[{{Baxter} {et~al.}(2015){Baxter}, {Keisler}, {Dodelson}, {Aird},
  {Allen}, {Ashby}, {Bautz}, {Bayliss}, {Benson}, {Bleem}, {Bocquet},
  {Brodwin}, {Carlstrom}, {Chang}, {Chiu}, {Cho}, {Clocchiatti}, {Crawford},
  {Crites}, {Desai}, {Dietrich}, {de Haan}, {Dobbs}, {Foley}, {Forman},
  {George}, {Gladders}, {Gonzalez}, {Halverson}, {Harrington}, {Hennig},
  {Hoekstra}, {Holder}, {Holzapfel}, {Hou}, {Hrubes}, {Jones}, {Knox}, {Lee},
  {Leitch}, {Liu}, {Lueker}, {Luong-Van}, {Mantz}, {Marrone}, {McDonald},
  {McMahon}, {Meyer}, {Millea}, {Mocanu}, {Murray}, {Padin}, {Pryke},
  {Reichardt}, {Rest}, {Ruhl}, {Saliwanchik}, {Saro}, {Sayre}, {Schaffer},
  {Shirokoff}, {Song}, {Spieler}, {Stalder}, {Stanford}, {Staniszewski},
  {Stark}, {Story}, {van Engelen}, {Vanderlinde}, {Vieira}, {Vikhlinin},
  {Williamson}, {Zahn}, \& {Zenteno}}]{Baxter2015}
{Baxter}, E.~J., {Keisler}, R., {Dodelson}, S., {et~al.} 2015, \apj, 806, 247,
  \dodoi{10.1088/0004-637X/806/2/247}

\bibitem[{{Chambers} {et~al.}(2016){Chambers}, {Magnier}, {Metcalfe},
  {Flewelling}, {Huber}, {Waters}, {Denneau}, {Draper}, {Farrow}, {Finkbeiner},
  {Holmberg}, {Koppenhoefer}, {Price}, {Rest}, {Saglia}, {Schlafly}, {Smartt},
  {Sweeney}, {Wainscoat}, {Burgett}, {Chastel}, {Grav}, {Heasley}, {Hodapp},
  {Jedicke}, {Kaiser}, {Kudritzki}, {Luppino}, {Lupton}, {Monet}, {Morgan},
  {Onaka}, {Shiao}, {Stubbs}, {Tonry}, {White}, {Ba{\~n}ados}, {Bell},
  {Bender}, {Bernard}, {Boegner}, {Boffi}, {Botticella}, {Calamida},
  {Casertano}, {Chen}, {Chen}, {Cole}, {Deacon}, {Frenk}, {Fitzsimmons},
  {Gezari}, {Gibbs}, {Goessl}, {Goggia}, {Gourgue}, {Goldman}, {Grant},
  {Grebel}, {Hambly}, {Hasinger}, {Heavens}, {Heckman}, {Henderson}, {Henning},
  {Holman}, {Hopp}, {Ip}, {Isani}, {Jackson}, {Keyes}, {Koekemoer}, {Kotak},
  {Le}, {Liska}, {Long}, {Lucey}, {Liu}, {Martin}, {Masci}, {McLean}, {Mindel},
  {Misra}, {Morganson}, {Murphy}, {Obaika}, {Narayan}, {Nieto-Santisteban},
  {Norberg}, {Peacock}, {Pier}, {Postman}, {Primak}, {Rae}, {Rai}, {Riess},
  {Riffeser}, {Rix}, {R{\"o}ser}, {Russel}, {Rutz}, {Schilbach}, {Schultz},
  {Scolnic}, {Strolger}, {Szalay}, {Seitz}, {Small}, {Smith}, {Soderblom},
  {Taylor}, {Thomson}, {Taylor}, {Thakar}, {Thiel}, {Thilker}, {Unger},
  {Urata}, {Valenti}, {Wagner}, {Walder}, {Walter}, {Watters}, {Werner},
  {Wood-Vasey}, \& {Wyse}}]{panstarrs}
{Chambers}, K.~C., {Magnier}, E.~A., {Metcalfe}, N., {et~al.} 2016, arXiv
  e-prints, arXiv:1612.05560.
\newblock \doarXiv{1612.05560}

\bibitem[{{Chisari} {et~al.}(2019){Chisari}, {Alonso}, {Krause}, {Leonard},
  {Bull}, {Neveu}, {Villarreal}, {Singh}, {McClintock}, {Ellison}, {Du},
  {Zuntz}, {Mead}, {Joudaki}, {Lorenz}, {Tr{\"o}ster}, {Sanchez}, {Lanusse},
  {Ishak}, {Hlozek}, {Blazek}, {Campagne}, {Almoubayyed}, {Eifler}, {Kirby},
  {Kirkby}, {Plaszczynski}, {Slosar}, {Vrastil}, {Wagoner}, \& {LSST Dark
  Energy Science Collaboration}}]{CCL}
{Chisari}, N.~E., {Alonso}, D., {Krause}, E., {et~al.} 2019, \apjs, 242, 2,
  \dodoi{10.3847/1538-4365/ab1658}

\bibitem[{{Chiu} {et~al.}(2020){Chiu}, {Umetsu}, {Murata}, {Medezinski}, \&
  {Oguri}}]{chiu19}
{Chiu}, I.~N., {Umetsu}, K., {Murata}, R., {Medezinski}, E., \& {Oguri}, M.
  2020, \mnras, 495, 428, \dodoi{10.1093/mnras/staa1158}

\bibitem[{{Di Mascolo} {et~al.}(2020){Di Mascolo}, {Mroczkowski}, {Churazov},
  {Moravec}, {Brodwin}, {Gonzalez}, {Decker}, {Eisenhardt}, {Stanford},
  {Stern}, {Sunyaev}, \& {Wylezalek}}]{madcows_aca}
{Di Mascolo}, L., {Mroczkowski}, T., {Churazov}, E., {et~al.} 2020, \aap, 638,
  A70, \dodoi{10.1051/0004-6361/202037818}

\bibitem[{{Dicker} {et~al.}(2020){Dicker}, {Romero}, {Di Mascolo},
  {Mroczkowski}, {Sievers}, {Moravec}, {Bhandarkar}, {Brodwin}, {Connor},
  {Decker}, {Devlin}, {Gonzalez}, {Lowe}, {Mason}, {Sarazin}, {Stanford},
  {Stern}, {Thongkham}, {Wylezalek}, \& {Zago}}]{madcows_mustang}
{Dicker}, S.~R., {Romero}, C.~E., {Di Mascolo}, L., {et~al.} 2020, arXiv
  e-prints, arXiv:2006.06703.
\newblock \doarXiv{2006.06703}

\bibitem[{{Diemer}(2018)}]{colossus}
{Diemer}, B. 2018, \apjs, 239, 35, \dodoi{10.3847/1538-4365/aaee8c}

\bibitem[{{Diemer} \& {Joyce}(2019)}]{diemer19}
{Diemer}, B., \& {Joyce}, M. 2019, \apj, 871, 168,
  \dodoi{10.3847/1538-4357/aafad6}

\bibitem[{{Dvornik} {et~al.}(2018){Dvornik}, {Hoekstra}, {Kuijken},
  {Schneider}, {Amon}, {Nakajima}, {Viola}, {Choi}, {Erben}, {Farrow},
  {Heymans}, {Hildebrand t}, {Sif{\'o}n}, \& {Wang}}]{dvornik18}
{Dvornik}, A., {Hoekstra}, H., {Kuijken}, K., {et~al.} 2018, \mnras, 479, 1240,
  \dodoi{10.1093/mnras/sty1502}

\bibitem[{Geach \& Peacock(2017)}]{GP2017}
Geach, J.~E., \& Peacock, J.~A. 2017, Nature Astronomy, 1, 795,
  \dodoi{10.1038/s41550-017-0259-1}

\bibitem[{{George} {et~al.}(2012){George}, {Leauthaud}, {Bundy}, {Finoguenov},
  {Ma}, {Rykoff}, {Tinker}, {Wechsler}, {Massey}, \& {Mei}}]{george12}
{George}, M.~R., {Leauthaud}, A., {Bundy}, K., {et~al.} 2012, \apj, 757, 2,
  \dodoi{10.1088/0004-637X/757/1/2}

\bibitem[{{Gonzalez} {et~al.}(2019){Gonzalez}, {Gettings}, {Brodwin},
  {Eisenhardt}, {Stanford}, {Wylezalek}, {Decker}, {Marrone}, {Moravec},
  {O'Donnell}, {Stalder}, {Stern}, {Abdulla}, {Brown}, {Carlstrom}, {Chambers},
  {Hayden}, {Lin}, {Magnier}, {Masci}, {Mantz}, {McDonald}, {Mo}, {Perlmutter},
  {Wright}, \& {Zeimann}}]{gonzalez19}
{Gonzalez}, A.~H., {Gettings}, D.~P., {Brodwin}, M., {et~al.} 2019, \apjs, 240,
  33, \dodoi{10.3847/1538-4365/aafad2}

\bibitem[{{G{\'o}rski} {et~al.}(2005){G{\'o}rski}, {Hivon}, {Banday},
  {Wandelt}, {Hansen}, {Reinecke}, \& {Bartelmann}}]{Healpix2}
{G{\'o}rski}, K.~M., {Hivon}, E., {Banday}, A.~J., {et~al.} 2005, \apj, 622,
  759, \dodoi{10.1086/427976}

\bibitem[{{Henderson} {et~al.}(2016){Henderson}, {Allison}, {Austermann},
  {Baildon}, {Battaglia}, {Beall}, {Becker}, {De Bernardis}, {Bond},
  {Calabrese}, {Choi}, {Coughlin}, {Crowley}, {Datta}, {Devlin}, {Duff},
  {Dunkley}, {D{\"u}nner}, {van Engelen}, {Gallardo}, {Grace}, {Hasselfield},
  {Hills}, {Hilton}, {Hincks}, {Hlo{\^z}ek}, {Ho}, {Hubmayr}, {Huffenberger},
  {Hughes}, {Irwin}, {Koopman}, {Kosowsky}, {Li}, {McMahon}, {Munson}, {Nati},
  {Newburgh}, {Niemack}, {Niraula}, {Page}, {Pappas}, {Salatino}, {Schillaci},
  {Schmitt}, {Sehgal}, {Sherwin}, {Sievers}, {Simon}, {Spergel}, {Staggs},
  {Stevens}, {Thornton}, {Van Lanen}, {Vavagiakis}, {Ward}, \&
  {Wollack}}]{Henderson2016}
{Henderson}, S.~W., {Allison}, R., {Austermann}, J., {et~al.} 2016, Journal of
  Low Temperature Physics, 184, 772, \dodoi{10.1007/s10909-016-1575-z}

\bibitem[{{Hilton} \& {ACT Collaboration}(2020)}]{Hilton2020}
{Hilton}, M., \& {ACT Collaboration}. 2020, in preparation

\bibitem[{{Hoekstra} {et~al.}(2013){Hoekstra}, {Bartelmann}, {Dahle}, {Israel},
  {Limousin}, \& {Meneghetti}}]{HoekstraReview}
{Hoekstra}, H., {Bartelmann}, M., {Dahle}, H., {et~al.} 2013, \ssr, 177, 75,
  \dodoi{10.1007/s11214-013-9978-5}

\bibitem[{{Hu} {et~al.}(2007){Hu}, {DeDeo}, \& {Vale}}]{astro-ph/0701276}
{Hu}, W., {DeDeo}, S., \& {Vale}, C. 2007, New Journal of Physics, 9, 441,
  \dodoi{10.1088/1367-2630/9/12/441}

\bibitem[{{Jee} {et~al.}(2011){Jee}, {Dawson}, {Hoekstra}, {Perlmutter},
  {Rosati}, {Brodwin}, {Suzuki}, {Koester}, {Postman}, {Lubin}, {Meyers},
  {Stanford}, {Barbary}, {Barrientos}, {Eisenhardt}, {Ford}, {Gilbank},
  {Gladders}, {Gonzalez}, {Harris}, {Huang}, {Lidman}, {Rykoff}, {Rubin}, \&
  {Spadafora}}]{jee11}
{Jee}, M.~J., {Dawson}, K.~S., {Hoekstra}, H., {et~al.} 2011, \apj, 737, 59,
  \dodoi{10.1088/0004-637X/737/2/59}

\bibitem[{{Johnston} {et~al.}(2007){Johnston}, {Sheldon}, {Wechsler}, {Rozo},
  {Koester}, {Frieman}, {McKay}, {Evrard}, {Becker}, \& {Annis}}]{johnston07}
{Johnston}, D.~E., {Sheldon}, E.~S., {Wechsler}, R.~H., {et~al.} 2007, arXiv
  e-prints, arXiv:0709.1159.
\newblock \doarXiv{0709.1159}

\bibitem[{{Madhavacheril} {et~al.}(2015){Madhavacheril}, {Sehgal}, {Allison},
  {Battaglia}, {Bond}, {Calabrese}, {Caliguiri}, {Coughlin}, {Crichton},
  {Datta}, {Devlin}, {Dunkley}, {D{\"u}nner}, {Fogarty}, {Grace}, {Hajian},
  {Hasselfield}, {Hill}, {Hilton}, {Hincks}, {Hlozek}, {Hughes}, {Kosowsky},
  {Louis}, {Lungu}, {McMahon}, {Moodley}, {Munson}, {Naess}, {Nati},
  {Newburgh}, {Niemack}, {Page}, {Partridge}, {Schmitt}, {Sherwin}, {Sievers},
  {Spergel}, {Staggs}, {Thornton}, {Van Engelen}, {Ward}, {Wollack}, \&
  {Atacama Cosmology Telescope Collaboration}}]{Mat2015}
{Madhavacheril}, M., {Sehgal}, N., {Allison}, R., {et~al.} 2015, Physical
  Review Letters, 114, 151302, \dodoi{10.1103/PhysRevLett.114.151302}

\bibitem[{{Madhavacheril} {et~al.}(2017){Madhavacheril}, {Battaglia}, \&
  {Miyatake}}]{madhavacheril17}
{Madhavacheril}, M.~S., {Battaglia}, N., \& {Miyatake}, H. 2017, \prd, 96,
  103525, \dodoi{10.1103/PhysRevD.96.103525}

\bibitem[{{Madhavacheril} \& {Hill}(2018)}]{1802.08230}
{Madhavacheril}, M.~S., \& {Hill}, J.~C. 2018, \prd, 98, 023534,
  \dodoi{10.1103/PhysRevD.98.023534}

\bibitem[{{Massey} {et~al.}(2010){Massey}, {Kitching}, \&
  {Richard}}]{MasseyReview}
{Massey}, R., {Kitching}, T., \& {Richard}, J. 2010, Reports on Progress in
  Physics, 73, 086901, \dodoi{10.1088/0034-4885/73/8/086901}

\bibitem[{{Miyatake} {et~al.}(2019){Miyatake}, {Battaglia}, {Hilton},
  {Medezinski}, {Nishizawa}, {More}, {Aiola}, {Bahcall}, {Bond}, {Calabrese},
  {Choi}, {Devlin}, {Dunkley}, {Dunner}, {Fuzia}, {Gallardo}, {Gralla},
  {Hasselfield}, {Halpern}, {Hikage}, {Hill}, {Hincks}, {Hlo{\v{z}}ek},
  {Huffenberger}, {Hughes}, {Koopman}, {Kosowsky}, {Louis}, {Madhavacheril},
  {McMahon}, {Mandelbaum}, {Marriage}, {Maurin}, {Miyazaki}, {Moodley},
  {Murata}, {Naess}, {Newburgh}, {Niemack}, {Nishimichi}, {Okabe}, {Oguri},
  {Osato}, {Page}, {Partridge}, {Robertson}, {Sehgal}, {Sherwin}, {Shirasaki},
  {Sievers}, {Sif{\'o}n}, {Simon}, {Spergel}, {Staggs}, {Stein}, {Takada},
  {Trac}, {Umetsu}, {van Engelen}, \& {Wollack}}]{miyatake19}
{Miyatake}, H., {Battaglia}, N., {Hilton}, M., {et~al.} 2019, \apj, 875, 63,
  \dodoi{10.3847/1538-4357/ab0af0}

\bibitem[{{Murata} {et~al.}(2019){Murata}, {Oguri}, {Nishimichi}, {Takada},
  {Mandelbaum}, {More}, {Shirasaki}, {Nishizawa}, \& {Osato}}]{murata19}
{Murata}, R., {Oguri}, M., {Nishimichi}, T., {et~al.} 2019, \pasj, 71, 107,
  \dodoi{10.1093/pasj/psz092}

\bibitem[{{Naess} {et~al.}(2020){Naess}, {Aiola}, {Austermann}, {Battaglia},
  {Beall}, {Becker}, {Bond}, {Calabrese}, {Choi}, {Cothard}, {Crowley},
  {Darwish}, {Datta}, {Denison}, {Devlin}, {Duell}, {Duff}, {Duivenvoorden},
  {Dunkley}, {D{\"u}nner}, {Fox}, {Gallardo}, {Halpern}, {Han}, {Hasselfield},
  {Hill}, {Hilton}, {Hilton}, {Hincks}, {Hlo{\v{z}}ek}, {Ho}, {Hubmayr},
  {Huffenberger}, {Hughes}, {Kosowsky}, {Louis}, {Madhavacheril}, {McMahon},
  {Moodley}, {Nati}, {Nibarger}, {Niemack}, {Page}, {Partridge}, {Salatino},
  {Schaan}, {Schillaci}, {Schmitt}, {Sherwin}, {Sehgal}, {Sif{\'o}n},
  {Spergel}, {Staggs}, {Stevens}, {Storer}, {Ullom}, {Vale}, {Van Engelen},
  {Van Lanen}, {Vavagiakis}, {Wollack}, \& {Xu}}]{2007.07290}
{Naess}, S., {Aiola}, S., {Austermann}, J.~E., {et~al.} 2020, arXiv e-prints,
  arXiv:2007.07290.
\newblock \doarXiv{2007.07290}

\bibitem[{{Navarro} {et~al.}(1996){Navarro}, {Frenk}, \& {White}}]{NFW}
{Navarro}, J.~F., {Frenk}, C.~S., \& {White}, S. D.~M. 1996, \apj, 462, 563,
  \dodoi{10.1086/177173}

\bibitem[{{Patil} {et~al.}(2020){Patil}, {Raghunathan}, \&
  {Reichardt}}]{1905.07943}
{Patil}, S., {Raghunathan}, S., \& {Reichardt}, C.~L. 2020, \apj, 888, 9,
  \dodoi{10.3847/1538-4357/ab55dd}

\bibitem[{{Peacock} \& {Smith}(2000)}]{peacock00}
{Peacock}, J.~A., \& {Smith}, R.~E. 2000, \mnras, 318, 1144,
  \dodoi{10.1046/j.1365-8711.2000.03779.x}

\bibitem[{{Planck Collaboration} \& {Ade}(2016)}]{PlnkSZCos2015}
{Planck Collaboration}, \& {Ade}, P.~A.~R. 2016, \aap, 594, A24,
  \dodoi{10.1051/0004-6361/201525833}

\bibitem[{{Planck Collaboration} {et~al.}(2016){Planck Collaboration}, {Ade},
  {Aghanim}, {Arnaud}, {Ashdown}, {Aumont}, {Baccigalupi}, {Banday},
  {Barreiro}, {Bartlett}, {Bartolo}, {Battaner}, {Battye}, {Benabed},
  {Beno{\^\i}t}, {Benoit-L{\'e}vy}, {Bernard}, {Bersanelli}, {Bielewicz},
  {Bock}, {Bonaldi}, {Bonavera}, {Bond}, {Borrill}, {Bouchet}, {Boulanger},
  {Bucher}, {Burigana}, {Butler}, {Calabrese}, {Cardoso}, {Catalano},
  {Challinor}, {Chamballu}, {Chary}, {Chiang}, {Chluba}, {Christensen},
  {Church}, {Clements}, {Colombi}, {Colombo}, {Combet}, {Coulais}, {Crill},
  {Curto}, {Cuttaia}, {Danese}, {Davies}, {Davis}, {de Bernardis}, {de Rosa},
  {de Zotti}, {Delabrouille}, {D{\'e}sert}, {Di Valentino}, {Dickinson},
  {Diego}, {Dolag}, {Dole}, {Donzelli}, {Dor{\'e}}, {Douspis}, {Ducout},
  {Dunkley}, {Dupac}, {Efstathiou}, {Elsner}, {En{\ss}lin}, {Eriksen},
  {Farhang}, {Fergusson}, {Finelli}, {Forni}, {Frailis}, {Fraisse},
  {Franceschi}, {Frejsel}, {Galeotta}, {Galli}, {Ganga}, {Gauthier}, {Gerbino},
  {Ghosh}, {Giard}, {Giraud-H{\'e}raud}, {Giusarma}, {Gjerl{\o}w},
  {Gonz{\'a}lez-Nuevo}, {G{\'o}rski}, {Gratton}, {Gregorio}, {Gruppuso},
  {Gudmundsson}, {Hamann}, {Hansen}, {Hanson}, {Harrison}, {Helou},
  {Henrot-Versill{\'e}}, {Hern{\'a}ndez-Monteagudo}, {Herranz}, {Hildebrand t},
  {Hivon}, {Hobson}, {Holmes}, {Hornstrup}, {Hovest}, {Huang}, {Huffenberger},
  {Hurier}, {Jaffe}, {Jaffe}, {Jones}, {Juvela}, {Keih{\"a}nen}, {Keskitalo},
  {Kisner}, {Kneissl}, {Knoche}, {Knox}, {Kunz}, {Kurki-Suonio}, {Lagache},
  {L{\"a}hteenm{\"a}ki}, {Lamarre}, {Lasenby}, {Lattanzi}, {Lawrence}, {Leahy},
  {Leonardi}, {Lesgourgues}, {Levrier}, {Lewis}, {Liguori}, {Lilje},
  {Linden-V{\o}rnle}, {L{\'o}pez-Caniego}, {Lubin}, {Mac{\'\i}as-P{\'e}rez},
  {Maggio}, {Maino}, {Mandolesi}, {Mangilli}, {Marchini}, {Maris}, {Martin},
  {Martinelli}, {Mart{\'\i}nez-Gonz{\'a}lez}, {Masi}, {Matarrese}, {McGehee},
  {Meinhold}, {Melchiorri}, {Melin}, {Mendes}, {Mennella}, {Migliaccio},
  {Millea}, {Mitra}, {Miville-Desch{\^e}nes}, {Moneti}, {Montier}, {Morgante},
  {Mortlock}, {Moss}, {Munshi}, {Murphy}, {Naselsky}, {Nati}, {Natoli},
  {Netterfield}, {N{\o}rgaard-Nielsen}, {Noviello}, {Novikov}, {Novikov},
  {Oxborrow}, {Paci}, {Pagano}, {Pajot}, {Paladini}, {Paoletti}, {Partridge},
  {Pasian}, {Patanchon}, {Pearson}, {Perdereau}, {Perotto}, {Perrotta},
  {Pettorino}, {Piacentini}, {Piat}, {Pierpaoli}, {Pietrobon}, {Plaszczynski},
  {Pointecouteau}, {Polenta}, {Popa}, {Pratt}, {Pr{\'e}zeau}, {Prunet},
  {Puget}, {Rachen}, {Reach}, {Rebolo}, {Reinecke}, {Remazeilles}, {Renault},
  {Renzi}, {Ristorcelli}, {Rocha}, {Rosset}, {Rossetti}, {Roudier},
  {Rouill{\'e} d'Orfeuil}, {Rowan-Robinson}, {Rubi{\~n}o-Mart{\'\i}n},
  {Rusholme}, {Said}, {Salvatelli}, {Salvati}, {Sandri}, {Santos},
  {Savelainen}, {Savini}, {Scott}, {Seiffert}, {Serra}, {Shellard}, {Spencer},
  {Spinelli}, {Stolyarov}, {Stompor}, {Sudiwala}, {Sunyaev}, {Sutton},
  {Suur-Uski}, {Sygnet}, {Tauber}, {Terenzi}, {Toffolatti}, {Tomasi},
  {Tristram}, {Trombetti}, {Tucci}, {Tuovinen}, {T{\"u}rler}, {Umana},
  {Valenziano}, {Valiviita}, {Van Tent}, {Vielva}, {Villa}, {Wade}, {Wandelt},
  {Wehus}, {White}, {White}, {Wilkinson}, {Yvon}, {Zacchei}, \&
  {Zonca}}]{Planck2015Cosmo}
{Planck Collaboration}, {Ade}, P.~A.~R., {Aghanim}, N., {et~al.} 2016, \aap,
  594, A13, \dodoi{10.1051/0004-6361/201525830}

\bibitem[{{Planck Collaboration} {et~al.}(2018){Planck Collaboration},
  {Akrami}, {Ashdown}, {Aumont}, {Baccigalupi}, {Ballardini}, {Band ay},
  {Barreiro}, {Bartolo}, {Basak}, {Benabed}, {Bersanelli}, {Bielewicz}, {Bond},
  {Borrill}, {Bouchet}, {Boulanger}, {Bucher}, {Burigana}, {Calabrese},
  {Cardoso}, {Carron}, {Casaponsa}, {Challinor}, {Colombo}, {Combet}, {Crill},
  {Cuttaia}, {de Bernardis}, {de Rosa}, {de Zotti}, {Delabrouille}, {Delouis},
  {Di Valentino}, {Dickinson}, {Diego}, {Donzelli}, {Dor{\'e}}, {Ducout},
  {Dupac}, {Efstathiou}, {Elsner}, {En{\ss}lin}, {Eriksen}, {Falgarone},
  {Fernandez-Cobos}, {Finelli}, {Forastieri}, {Frailis}, {Fraisse},
  {Franceschi}, {Frolov}, {Galeotta}, {Galli}, {Ganga}, {G{\'e}nova-Santos},
  {Gerbino}, {Ghosh}, {Gonz{\'a}lez-Nuevo}, {G{\'o}rski}, {Gratton},
  {Gruppuso}, {Gudmundsson}, {Hand ley}, {Hansen}, {Helou}, {Herranz}, {Huang},
  {Jaffe}, {Karakci}, {Keih{\"a}nen}, {Keskitalo}, {Kiiveri}, {Kim}, {Kisner},
  {Krachmalnicoff}, {Kunz}, {Kurki-Suonio}, {Lagache}, {Lamarre}, {Lasenby},
  {Lattanzi}, {Lawrence}, {Le Jeune}, {Levrier}, {Liguori}, {Lilje},
  {Lindholm}, {L{\'o}pez-Caniego}, {Lubin}, {Ma}, {Mac{\'\i}as-P{\'e}rez},
  {Maggio}, {Maino}, {Mandolesi}, {Mangilli}, {Marcos-Caballero}, {Martin},
  {Mart{\'\i}nez-Gonz{\'a}lez}, {Matarrese}, {Mauri}, {McEwen}, {Meinhold},
  {Melchiorri}, {Mennella}, {Migliaccio}, {Miville-Desch{\^e}nes}, {Molinari},
  {Moneti}, {Montier}, {Morgante}, {Natoli}, {Oppizzi}, {Pagano}, {Paoletti},
  {Partridge}, {Peel}, {Pettorino}, {Piacentini}, {Polenta}, {Puget}, {Rachen},
  {Reinecke}, {Remazeilles}, {Renzi}, {Rocha}, {Roudier},
  {Rubi{\~n}o-Mart{\'\i}n}, {Ruiz-Granados}, {Salvati}, {Sandri}, {Savelainen},
  {Scott}, {Seljebotn}, {Sirignano}, {Spencer}, {Suur-Uski}, {Tauber},
  {Tavagnacco}, {Tenti}, {Thommesen}, {Toffolatti}, {Tomasi}, {Trombetti},
  {Valiviita}, {Van Tent}, {Vielva}, {Villa}, {Vittorio}, {Wandelt}, {Wehus},
  {Zacchei}, \& {Zonca}}]{Planck2018compsep}
{Planck Collaboration}, {Akrami}, Y., {Ashdown}, M., {et~al.} 2018, arXiv
  e-prints, arXiv:1807.06208.
\newblock \doarXiv{1807.06208}

\bibitem[{{Raghunathan} {et~al.}(2019){Raghunathan}, {Patil}, {Baxter},
  {Benson}, {Bleem}, {Chou}, {Crawford}, {Holder}, {McClintock}, {Reichardt},
  {Rozo}, {Varga}, {Abbott}, {Ade}, {Allam}, {Anderson}, {Annis}, {Austermann},
  {Avila}, {Beall}, {Bechtol}, {Bender}, {Bernstein}, {Bertin}, {Bianchini},
  {Brooks}, {Burke}, {Carlstrom}, {Carretero}, {Chang}, {Chiang}, {Cho},
  {Citron}, {Crites}, {Cunha}, {da Costa}, {Davis}, {Desai}, {Diehl},
  {Dietrich}, {Dobbs}, {Doel}, {Eifler}, {Everett}, {Evrard}, {Flaugher},
  {Fosalba}, {Frieman}, {Gallicchio}, {Garc{\'\i}a-Bellido}, {Gaztanaga},
  {George}, {Gilbert}, {Gruen}, {Gruendl}, {Gschwend}, {Gupta}, {Gutierrez},
  {de Haan}, {Halverson}, {Harrington}, {Hartley}, {Henning}, {Hilton},
  {Hollowood}, {Holzapfel}, {Honscheid}, {Hou}, {Hoyle}, {Hrubes}, {Huang},
  {Hubmayr}, {Irwin}, {James}, {Jeltema}, {Kim}, {Carrasco Kind}, {Knox},
  {Kovacs}, {Kuehn}, {Kuropatkin}, {Lee}, {Li}, {Lima}, {Maia}, {Marshall},
  {McMahon}, {Melchior}, {Menanteau}, {Meyer}, {Miller}, {Miquel}, {Mocanu},
  {Montgomery}, {Nadolski}, {Natoli}, {Nibarger}, {Novosad}, {Padin}, {Plazas},
  {Pryke}, {Rapetti}, {Romer}, {Carnero Rosell}, {Ruhl}, {Saliwanchik},
  {Sanchez}, {Sayre}, {Scarpine}, {Schaffer}, {Schubnell}, {Serrano},
  {Sevilla-Noarbe}, {Smecher}, {Smith}, {Soares-Santos}, {Sobreira}, {Stark},
  {Story}, {Suchyta}, {Swanson}, {Tarle}, {Thomas}, {Tucker}, {Vanderlinde},
  {De Vicente}, {Vieira}, {Wang}, {Whitehorn}, {Wu}, \&
  {Zhang}}]{2019ApJ...872..170R}
{Raghunathan}, S., {Patil}, S., {Baxter}, E., {et~al.} 2019, \apj, 872, 170,
  \dodoi{10.3847/1538-4357/ab01ca}

\bibitem[{{Rozo} {et~al.}(2014){Rozo}, {Bartlett}, {Evrard}, \&
  {Rykoff}}]{rozo14}
{Rozo}, E., {Bartlett}, J.~G., {Evrard}, A.~E., \& {Rykoff}, E.~S. 2014,
  \mnras, 438, 78, \dodoi{10.1093/mnras/stt2161}

\bibitem[{{Schrabback} {et~al.}(2018){Schrabback}, {Applegate}, {Dietrich},
  {Hoekstra}, {Bocquet}, {Gonzalez}, {von der Linden}, {McDonald}, {Morrison},
  {Raihan}, {Allen}, {Bayliss}, {Benson}, {Bleem}, {Chiu}, {Desai}, {Foley},
  {de Haan}, {High}, {Hilbert}, {Mantz}, {Massey}, {Mohr}, {Reichardt}, {Saro},
  {Simon}, {Stern}, {Stubbs}, \& {Zenteno}}]{schrabback18}
{Schrabback}, T., {Applegate}, D., {Dietrich}, J.~P., {et~al.} 2018, \mnras,
  474, 2635, \dodoi{10.1093/mnras/stx2666}

\bibitem[{{Seljak}(2000)}]{seljak00}
{Seljak}, U. 2000, \mnras, 318, 203, \dodoi{10.1046/j.1365-8711.2000.03715.x}

\bibitem[{{Swetz} {et~al.}(2011){Swetz}, {Ade}, {Amiri}, {Appel},
  {Battistelli}, {Burger}, {Chervenak}, {Devlin}, {Dicker}, {Doriese},
  {D{\"u}nner}, {Essinger-Hileman}, {Fisher}, {Fowler}, {Halpern},
  {Hasselfield}, {Hilton}, {Hincks}, {Irwin}, {Jarosik}, {Kaul}, {Klein},
  {Lau}, {Limon}, {Marriage}, {Marsden}, {Martocci}, {Mauskopf}, {Moseley},
  {Netterfield}, {Niemack}, {Nolta}, {Page}, {Parker}, {Staggs}, {Stryzak},
  {Switzer}, {Thornton}, {Tucker}, {Wollack}, \& {Zhao}}]{Swetz2011}
{Swetz}, D.~S., {Ade}, P.~A.~R., {Amiri}, M., {et~al.} 2011, \apjs, 194, 41,
  \dodoi{10.1088/0067-0049/194/2/41}

\bibitem[{{Thornton} {et~al.}(2016){Thornton}, {Ade}, {Aiola}, {Angil{\`e}},
  {Amiri}, {Beall}, {Becker}, {Cho}, {Choi}, {Corlies}, {Coughlin}, {Datta},
  {Devlin}, {Dicker}, {D{\"u}nner}, {Fowler}, {Fox}, {Gallardo}, {Gao},
  {Grace}, {Halpern}, {Hasselfield}, {Henderson}, {Hilton}, {Hincks}, {Ho},
  {Hubmayr}, {Irwin}, {Klein}, {Koopman}, {Li}, {Louis}, {Lungu}, {Maurin},
  {McMahon}, {Munson}, {Naess}, {Nati}, {Newburgh}, {Nibarger}, {Niemack},
  {Niraula}, {Nolta}, {Page}, {Pappas}, {Schillaci}, {Schmitt}, {Sehgal},
  {Sievers}, {Simon}, {Staggs}, {Tucker}, {Uehara}, {van Lanen}, {Ward}, \&
  {Wollack}}]{thornton/2016}
{Thornton}, R.~J., {Ade}, P.~A.~R., {Aiola}, S., {et~al.} 2016, \apjs, 227, 21,
  \dodoi{10.3847/1538-4365/227/2/21}

\bibitem[{{Tinker} {et~al.}(2010){Tinker}, {Robertson}, {Kravtsov}, {Klypin},
  {Warren}, {Yepes}, \& {Gottl{\"o}ber}}]{tinker10}
{Tinker}, J.~L., {Robertson}, B.~E., {Kravtsov}, A.~V., {et~al.} 2010, \apj,
  724, 878, \dodoi{10.1088/0004-637X/724/2/878}

\bibitem[{{Trimble}(1987)}]{TrimbleReview}
{Trimble}, V. 1987, \araa, 25, 425, \dodoi{10.1146/annurev.aa.25.090187.002233}

\bibitem[{{van den Bosch} {et~al.}(2013){van den Bosch}, {More}, {Cacciato},
  {Mo}, \& {Yang}}]{vdbosch13}
{van den Bosch}, F.~C., {More}, S., {Cacciato}, M., {Mo}, H., \& {Yang}, X.
  2013, \mnras, 430, 725, \dodoi{10.1093/mnras/sts006}

\bibitem[{{Viola} {et~al.}(2015){Viola}, {Cacciato}, {Brouwer}, {Kuijken},
  {Hoekstra}, {Norberg}, {Robotham}, {van Uitert}, {Alpaslan}, {Baldry},
  {Choi}, {de Jong}, {Driver}, {Erben}, {Grado}, {Graham}, {Heymans},
  {Hildebrand t}, {Hopkins}, {Irisarri}, {Joachimi}, {Loveday}, {Miller},
  {Nakajima}, {Schneider}, {Sif{\'o}n}, \& {Verdoes Kleijn}}]{viola15}
{Viola}, M., {Cacciato}, M., {Brouwer}, M., {et~al.} 2015, \mnras, 452, 3529,
  \dodoi{10.1093/mnras/stv1447}

\bibitem[{{Voit}(2005)}]{Voit2005}
{Voit}, G.~M. 2005, Reviews of Modern Physics, 77, 207,
  \dodoi{10.1103/RevModPhys.77.207}

\bibitem[{{Wright} \& {Brainerd}(2000)}]{wright00}
{Wright}, C.~O., \& {Brainerd}, T.~G. 2000, \apj, 534, 34,
  \dodoi{10.1086/308744}

\bibitem[{{Wright} {et~al.}(2010){Wright}, {Eisenhardt}, {Mainzer}, {Ressler},
  {Cutri}, {Jarrett}, {Kirkpatrick}, {Padgett}, {McMillan}, {Skrutskie},
  {Stanford}, {Cohen}, {Walker}, {Mather}, {Leisawitz}, {Gautier}, {McLean},
  {Benford}, {Lonsdale}, {Blain}, {Mendez}, {Irace}, {Duval}, {Liu}, {Royer},
  {Heinrichsen}, {Howard}, {Shannon}, {Kendall}, {Walsh}, {Larsen}, {Cardon},
  {Schick}, {Schwalm}, {Abid}, {Fabinsky}, {Naes}, \& {Tsai}}]{WISE2010}
{Wright}, E.~L., {Eisenhardt}, P.~R.~M., {Mainzer}, A.~K., {et~al.} 2010, \aj,
  140, 1868, \dodoi{10.1088/0004-6256/140/6/1868}

\bibitem[{{Zonca} {et~al.}(2019){Zonca}, {Singer}, {Lenz}, {Reinecke},
  {Rosset}, {Hivon}, \& {Gorski}}]{Healpix1}
{Zonca}, A., {Singer}, L., {Lenz}, D., {et~al.} 2019, The Journal of Open
  Source Software, 4, 1298, \dodoi{10.21105/joss.01298}

\bibitem[{{Zubeldia} \& {Challinor}(2019)}]{2019MNRAS.489..401Z}
{Zubeldia}, {\'I}., \& {Challinor}, A. 2019, \mnras, 489, 401,
  \dodoi{10.1093/mnras/stz2153}

\end{thebibliography}
\bibliographystyle{aasjournal}

\appendix

\section{Method details and systematics tests}
\label{app:details}

The quadratic estimator for CMB lensing reconstruction can be recast as the divergence of the product of the small-scale CMB and the large-scale CMB gradient:

\begin{equation}
\label{Eq:kappaEst}
\hat{\kappa}(\boldsymbol{\theta}) =-\mathcal{F}^{-1}\left\{A^{TT}(\bL) \mathcal{F}\left\{ \rmn{Re}\left[\nabla \cdot \left[\boldsymbol{\nabla}T_g(\boldsymbol{\theta}) T_h(\boldsymbol{\theta}) \right]\right]\right\}\right\}
\end{equation}
where $\mathcal{F}$ and $\mathcal{F}^{-1}$ denote 2d Fourier and inverse Fourier transforms respectively. Our analysis applies this estimator locally to cut-outs of CMB data centered on each cluster location that are 128 arcminutes wide  with pixels of width 0.5 arcminutes. The large width relative to the typical arcminute size of clusters allows us to estimate the large-scale gradient.

Following \cite{astro-ph/0701276}, we use a low-pass (top-hat) filtered temperature anisotropy gradient $\boldsymbol{\nabla}T_g$ with a maximum CMB multipole of $\ell_G=2000$ to mitigate bias in massive clusters as well as to reduce contamination from foregrounds. Furthermore, following \cite{1802.08230}, for the gradient map $\boldsymbol{\nabla}T_g$, we use a CMB map from which tSZ has been explicitly deprojected using multi-frequency information (the \Planck\ PR3 SMICA tSZ-deprojected), so as to null a large tSZ-induced bias. For the high-resolution map $T_h$, we use an internal linear combination (ILC) of the postage stamp cut-outs from the 98 GHz and 150 GHz 2018 co-adds of \Planck\ and ACT from \cite{2007.07290}, where the high-resolution ACT data dominates the information content. While the tSZ cleaning for $\boldsymbol{\nabla}T_g$ nulls the tSZ bias, a large amount of variance can be induced by the presence of large tSZ decrements in $T_h$. To reduce this, following \cite{1905.07943}, prior to the ILC we subtract best estimates of the tSZ decrements from each of the $\sim 4000$ SZ clusters detected in the co-add \citep{Hilton2020} since some of these clusters either appear in our sample or may appear in the postage stamps that we perform our reconstruction on. This has the effect of reducing the uncertainty in the first three bins of our measurement by $\approx 10\%$. The weights for the ILC are designed to minimize the power spectrum of the combination and are determined after fitting the total 1d power spectrum of each single-frequency stamp to a simple two parameter model that captures both an atmospheric component as well as white noise. We impose a maximum multipole cut of $\ell=6000$ on the high-resolution map.  Both maps are inverse-variance filtered as in \cite{astro-ph/0701276}. For both the gradient and high-resolution map, we do not use scales below $\ell=200$ due to the size of our cut-outs.
The reconstruction is normalized with an analytic expression $A^{TT}(\bL)$ \citep{astro-ph/0701276} that depends on the filters we apply. The final reconstruction is filtered to ensure it only contains modes $200 < L < 5000$, and this filter is subsequently propagated in our theoretical model when fitting the profile. 

Prior to reconstruction, the gradient and high-resolution maps are multiplied by a tapering cosine window (of approximate width 18 arcminutes) to enforce periodicity required by Fourier transforms. The presence of such a window (as well as other sources of anisotropy such as inhomogeneity of instrument noise) induces a spurious lensing signal referred to as the `mean-field'. We estimate the mean-field by repeating the above stacking procedure on a large number of random locations ($\sim 300$ times the number of clusters in our sample) and subtract this from our main cluster stack. The mean-field profile is roughly constant as a function of distance from the center of the stack and is 20\% of the measurement in the first bin, 50\% in the second bin and larger than the measurement in subsequent bins.

We perform a number of systematics tests to ensure the robustness of our detection. This includes a curl null test, where the divergence in Eq. \ref{Eq:kappaEst} is replaced by the curl (and normalized appropriately), shown in Fig. \ref{fig:profile}. With this test, we obtain a measurement that is consistent with null (PTE of \curlpte, corresponding to consistency with null at the $\curlsigma\sigma$ level). As shown in \cite{1802.08230}, one needs correlated contaminants in the gradient and high-resolution maps of the quadratic estimator in order to be biased by these contaminants. Therefore, to test the possibility that dust emission might be biasing our measurement, we stack the \Planck\ PR3 SMICA tSZ-deprojected map (used for the gradient) at the location of all \tnclusters MaDCoWS clusters with photometric redshifts (not just the \nclusters that fall in the ACT footprint used in this analysis). We detect no residual in this stack and thus conclude that dust contamination in the lensing reconstruction itself is unlikely. The presence of contaminants in the high-resolution map will contribute noise, which is captured in our empirically determined covariance matrix.

\section{Mass Modeling}\label{app:model}

In order to infer the mean mass of the sample, we model the CMB lensing signal by assuming that all mass is contained within spherical halos, and that these halos cluster together \citep[i.e., using the `halo model';][]{peacock00,seljak00}. We adopt the best-fit flat $\Lambda$CDM cosmology from \cite{Planck2015Cosmo}, with present-day matter density parameter $\Omega_\mathrm{m}=0.307$ and a Hubble constant $H_0=67.7\,\mathrm{km\,s^{-1}Mpc^{-1}}$, also adopted by \cite{gonzalez19}.

Within the halo model formalism, the lensing signal can be decomposed as
\begin{equation}\label{eq:kappa_tot}
    \kappa(\theta) = 
        \kappa_\mathrm{1h}(\theta) + \kappa_\mathrm{2h}(\theta)
\end{equation}
where the subscripts 1h and 2h represent the 1-halo (due to each halo) and 2-halo (due to halo clustering) contributions. We model each component as follows.

For the 1-halo term, we use a Navarro-Frenk-White (NFW) density profile \citep{NFW}. We use a mass definition, $M_{500c}$, corresponding to the mass within an overdensity of 500 times the critical density, $\rho_\mathrm{c}(z)=3H^2(z)/8\pi G$.
We model the concentration, $c_{500c}\equiv r_{500c}/r_\mathrm{s}$ (where $r_\mathrm{s}$ is the scale radius and $r_{500c}$ is the radius containing $M_{500c}$), using the relation between concentration and mass from \cite{diemer19}.\footnote{Calculated using \texttt{colossus} \citep[\url{https://bdiemer.bitbucket.io/colossus/}]{colossus}.} For a lens of mass $m$ and redshift $z$, the convergence is related to the mass surface density, $\Sigma(m,z)$, through
\begin{equation}\label{eq:kappa}
    \kappa(m,z)=\frac{\Sigma(m,z)}{\Sigma_\mathrm{cr}(z)}
    \equiv
    \left(
        \frac{c^2}{4\pi G}\frac{D_\mathrm{s}}{D_\mathrm{l}(z)D_\mathrm{ls}(z)}
    \right)^{-1}
    \Sigma(m,z),
\end{equation}
where $D_\mathrm{s}$, $D_\mathrm{l}$, and $D_\mathrm{ls}$ are the angular diameter distances to the last scattering surface, to the cluster, and between the cluster and the last scattering surface, respectively.
The expression for the NFW surface density profile is given in \cite{wright00}.

The 2-halo term arises due to the large-scale galaxy-matter power spectrum,
\begin{equation}\label{eq:Pgm_2h}
    P_\mathrm{gm}^\mathrm{2h}(k|m,z) = b(m,z)P_\mathrm{m}(k|z)
\end{equation}
where $P_\mathrm{m}(k|z)$ is the linear matter power spectrum and $b(m,z)$ is the halo bias as calculated by \cite{tinker10}. Note that all calculations of the 2-halo term are performed in comoving coordinates. The power spectrum is the Fourier transform of the correlation function,
\begin{equation}\label{eq:xi}
    \xi_\mathrm{gm}(r|m,z) = 
    \frac1{2\pi^2}\int_0^\infty \frac{\sin kr}{kr}k^2\,dk\, P_\mathrm{gm}^\mathrm{2h}
\end{equation}
from which the surface density can be calculated as
\begin{equation}\label{eq:sigma}
    \Sigma_\mathrm{2h}(R|m,z) = 2\bar\rho_\mathrm{m}R
    \int_0^1\frac{dx}{x^2\sqrt{1-x^2}}\,\xi_\mathrm{gm}(R/x|m,z)
\end{equation}
where $\bar\rho_\mathrm{m}$ is the (time-independent) comoving mean matter density. The 2-halo term in Eq. \ref{eq:kappa_tot} is then
\begin{equation}
    \kappa_\mathrm{2h}(\theta|m,z) = (1+z)^2
        \frac{\Sigma_\mathrm{2h}(\theta|m,z)}{\Sigma_\mathrm{c}}
\end{equation}
where the additional $(1+z)^2$ compared to Eq. \ref{eq:kappa} is due to our use of comoving coordinates in the 2-halo term calculation \citep[see, e.g., ][]{dvornik18}.
We then filter both the 1h and 2h components with the Fourier-space filter described in Appendix \ref{app:details} to produce a prediction for $\hat\kappa(\theta|m,z)$.

In reality, our measurement is the mean over clusters covering a range in masses and redshifts. The halo mass function provides a prediction for the distribution of clusters in mass and redshift, which can be linked to the observable used to construct the sample (namely, richness, $\observable$) through a mass-observable relation. However, like for any other cluster sample, due to the selection algorithm (in this case, red sequence overdensities), \madcows\ is a biased subsample of the underlying cluster population. We therefore follow previous galaxy-galaxy- and cluster-lensing studies \citep[e.g.,][]{vdbosch13,viola15,dvornik18,miyatake19} and model the measured lensing signal as
\begin{equation}\label{eq:kappa_avg}
% \begin{split}
    \hat\kappa(\theta) = \frac1{\bar N}\int dz \,{\frac{dV_\mathrm{eff}}{dz}}\int dm \,{\frac{dn_\mathrm{h}}{dm}} %\\
    \times
    \int d\observable \mathcal{P}(\ln\observable|\ln m,z)\mathcal{S}(\observable,z)\,\hat\kappa(\theta|m,z),
% \end{split}
\end{equation}
 {Here $\frac{dV_\mathrm{eff}}{dz}=A\frac{dN_\mathrm{cl}}{dz}$ is the effective volume probed per unit redshift, which accounts for the unknown redshift component in the selection function (and $A$ is a constant that absorbs the unit conversion, but cancels between the numerator and denominator);
$\frac{dn_\mathrm{h}}{dm}$} is the halo mass function from \cite{tinker10}\footnote{We use the Core Cosmology Library \citep[\url{https://ccl.readthedocs.io/en/latest/}]{CCL} to calculate the halo mass function, the halo bias, and the matter power spectrum.}, $\mathcal{P}(\ln\observable|\ln m,z)$ is the conditional probability of richness given cluster mass (defined below), and $\bar N$ is the expected number of clusters, which differs from the halo mass function due to the mass-observable relation (see below) and selection function, $\mathcal{S}(\observable,z)$, which is given by
\begin{equation}\label{eq:nbar}
% \begin{split}
    \bar N = \int dz \,{\frac{dV_\mathrm{eff}}{dz}}\int dm \,{\frac{dn_\mathrm{h}}{dm}} %\\
    \times
    \int d\observable \mathcal{P}(\ln\observable|\ln m,z)\mathcal{S}(\observable,z).
% \end{split}
\end{equation}
Consequently, in this framework mean values refer to the mean of the underlying cluster population and are calculated in analogy to eq.~\ref{eq:kappa_avg} as
\begin{equation}\label{eq:avg}
% \begin{split}
    \langle X\rangle = \frac1{\bar N}\int dz \, {\frac{dV_\mathrm{eff}}{dz}}\int dm \,{\frac{dn_\mathrm{h}}{dm}} %\\
    \times
    \int d\observable \mathcal{P}(\ln\observable|\ln m,z)\mathcal{S}(\observable,z) \,X.
% \end{split}
\end{equation}
For simplicity, we assume that the \madcows\ sample includes all clusters with richness larger than 20 (and we have discarded all clusters with lower richness), such that
\begin{equation}
    \mathcal{S}(\observable,z) = \mathcal{S}(\observable) = \Theta(\observable-20),
\end{equation}
where $\Theta$ is the Heaviside step function. 
In addition, we limit the redshift integral to the range $z\in[0.7,1.8]$. (Note that by using the observed redshift distribution, $\frac{dN_\mathrm{cl}}{dz}$, instead of the comoving volume element $dV/dz$ in Eq.\ \ref{eq:kappa_avg}, we are effectively introducing an additional redshift component to the selection function; this choice has no impact on our results.)

Additionally, we assume a linear relation between cluster richness, $\observable$, and cluster mass, $M_{500c}\equiv m$:
\begin{equation}\label{eq:scaling}
\ln \observable = \beta + \ln (m/10^{14}\,\Msun) \equiv \beta + \ln\mu
\end{equation}
and assume a log-normal conditional probability of $\observable$ given cluster mass,
\begin{equation}
    \mathcal{P}(\ln\observable|\ln m,z) = \frac1{\sqrt{2\pi}\sigma}
    \exp\left[-\frac{(\ln\observable-\ln\mu-\beta)^2}{2\sigma^2}\right].
\end{equation}
We assume a scatter on the mass-observable relation $\sigma\equiv\sigma_{\ln\observable|\ln m}=0.5$, typical of the scatter of different richness definitions \citep{rozo14,murata19}. The normalization $\beta$ is the only free parameter in this model. While its posterior value depends on our assumptions about the selection function and the scatter above, the mean mass of the sample (calculated following eq.\ \ref{eq:avg}) is robust to these changes as it directly quantifies the amplitude of the lensing signal.

We estimate the best-fit parameters by varying the normalization $\beta$ and minimizing
\begin{equation}
    \chi^2 = \left(\mathbf{d-m}\right)^T\cdot\mathbf{C}^{-1}\cdot\left(\mathbf{d-m}\right)
\end{equation}
where $\mathbf{d}$ and $\mathbf{m}$ are the vectors of measured radial profile bins and the model expectations for them, respectively, and $\mathbf{C}$ is the covariance matrix of the measurement (see Eq. \ref{eq:cov}). We use five bins in our fit that encompass a region within 8 arcminutes of the center of the stamp.

In order to demonstrate the contribution from each term in the model described above, we first fit our measurements with a single NFW profile at $z=1.1$, including only the 1-halo contribution, with the mass as the sole free parameter. This results in a best-fit mass $\langle M_{500c}\rangle=\left(2.1\pm0.5\right)\times10^{14}\,\Msun$. We then add the two-halo term, assuming all clusters have the same mass and are located at the same mean redshift, and find $\langle M_{500c}\rangle=\left(1.7\pm0.4\right)\times10^{14}\,\Msun$.
Finally, we implement the full model as described above. We find $\beta=2.4\pm0.3$, which translates to a population-weighted mean cluster mass $\langle M_{500c}\rangle = \left(\meanmass\right)\times10^{14}\,\Msun$, where uncertainties correspond to $\Delta\chi^2=1$ (i.e., 68.3\% credible range). This is the main result of this paper.

For reference, we calculate the mass $M_{200c}$ within an overdensity of 200 times the critical density for each point in the $(M_{500c},z)$ grid and perform the integral given by Eq.\ \ref{eq:avg} over the appropriate halo mass function, and find $\langle M_{200c}\rangle=(2.5\pm0.6)\times10^{14}\,\Msun$.

As discussed above, the best-fit $\beta$ depends strongly on our modeling assumptions and should not be over-interpreted, but the mean mass changes at most by 5\% regardless of our assumptions about the selection function and scaling relation. Specifically, we have varied the scaling relation normalization, $\beta$, and intrinsic scatter, $\sigma$, by factors of two from the adopted values, as well as selection functions given both by step functions and error functions with mid-points in the range $\lambda=[10,30]$, and in all cases find mean masses within 5\% of the reported value. The largest shift in mass is introduced by modifying the mass-concentration relation (for reference, the population-weighted mean concentration is $c_{500c}=2.5$): increasing (decreasing) the concentration by 20\%---which brackets most mass-concentration relations in the literature---increases (decreases) the mean mass by 10\%, well within the uncertainties. Redshift uncertainties also do not change our results significantly: applying a systematic shift $\Delta z=\pm0.1$ (slighly larger than the 1$\sigma$ scatter of photometric redshifts compared to spectroscopic redshifts, \cite{gonzalez19}) only changes the mean mass by less than 3\%. Assuming that 5\% of the clusters have photometric redshifts that are wrong by $\Delta z=+0.2$ ($3\sigma$ outliers, all in the same direction) has a similar impact. 

Our final assessment of modeling systematics regards miscentering.
The \madcows\ algorithm defines the cluster center as the location of the peak amplitude in a smoothed galaxy density map, which can potentially be significantly offset from the center of mass \citep{george12,viola15} and results in a reduced central lensing amplitude.
However, the physical resolution of our measurements is roughly 1 Mpc, which greatly reduces the impact from miscentering, and we choose not to account for it in our analysis. To test the impact of miscentering, we implement the distribution found by \cite{johnston07} for optically-selected clusters, exploring typical offsets between the assumed and true centers of up to 0.4 Mpc (roughly $2r_\mathrm{s}$); the change in mass is at most 5\%.

\end{document}